\documentclass[aps,prd,twocolumn,preprintnumbers,nofootinbib,amsmath,amssymb,superscriptaddress,10pt,floatfix]{revtex4-2}

\usepackage{graphicx}
\usepackage{placeins}

\usepackage[hidelinks]{hyperref}
\usepackage[capitalise]{cleveref}
\crefname{section}{Sec.}{Sections} 

\usepackage[font=small, labelfont=bf, textfont={small,it}]{caption}
\usepackage{multirow}
\usepackage{makecell}
\usepackage{enumitem}

\usepackage{xcolor}
\usepackage{xspace}
\usepackage[normalem]{ulem}
\usepackage{braket}

\newcommand{\bee}{\begin{equation}}
\newcommand{\eee}{\end{equation}}
\newcommand{\beq}{\begin{eqnarray}}
\newcommand{\eeq}{\end{eqnarray}}
\newcommand{\beqnn}{\begin{eqnarray*}}
\newcommand{\eeqnn}{\end{eqnarray*}}

\newcommand{\sst}[1]{\scriptscriptstyle{#1}}
\newcommand{\ii}{\mathrm{i}}
\newcommand{\ee}{\mathrm{e}}
\newcommand{\dd}{\mathrm{d}}
\newcommand{\Tr}{\mathrm{Tr}}
\renewcommand{\S}{{\sst{\rm S}}}
\renewcommand{\top}{{\sst{\rm top}}}
\newcommand{\QCD}{{\sst{\rm QCD}}}
\newcommand{\E}{{\sst{\rm E}}}
\newcommand{\LL}{{\sst{\rm L}}}
\newcommand{\Rs}{R_{\sst{\rm s}}}
\newcommand{\ncool}{n_{\sst{\rm cool}}}
\newcommand{\rhobar}{\bar{\rho}}
\newcommand{\omegastar}{\omega^{*}}
\newcommand{\alphas}{\alpha_{s}}
\newcommand{\mD}{m_{\sst{\rm D}}}

\begin{document}

\title{Real-time topological rate at non-zero momentum in quenched QCD}

\author{Claudio Bonanno}
\email{claudio.bonanno@unibe.ch}
\affiliation{Albert Einstein Center for Fundamental Physics, Institute for Theoretical Physics, University of Bern, Sidlerstra{\ss}e 5, CH-3012 Bern, Switzerland}

\author{Massimo D'Elia}
\email{massimo.delia@unipi.it}
\affiliation{Dipartimento di Fisica, Università di Pisa \& INFN, Sezione di Pisa, Largo Pontecorvo 3, I-56127 Pisa, Italy}

\author{Roberto Dionisio}
\email{roberto.dionisio@phd.unipi.it}
\affiliation{Dipartimento di Fisica, Università di Pisa \& INFN, Sezione di Pisa, Largo Pontecorvo 3, I-56127 Pisa, Italy}

\author{Giuseppe Gagliardi}
\email{giuseppe.gagliardi@roma3.infn.it}
\affiliation{INFN, Sezione di Roma Tre, Via della Vasca Navale 84, I-00146 Rome, Italy}

\author{Andrea Giorgieri}
\email{andrea.giorgieri@phd.unipi.it}
\affiliation{Dipartimento di Fisica, Università di Pisa \& INFN, Sezione di Pisa, Largo Pontecorvo 3, I-56127 Pisa, Italy}

\author{Francesco Sanfilippo}
\email{francesco.sanfilippo@roma3.infn.it}
\affiliation{INFN, Sezione di Roma Tre, Via della Vasca Navale 84, I-00146 Rome, Italy}

\author{Alessandra Valentino}
\email{a.valentino4@studenti.unipi.it}
\affiliation{Dipartimento di Fisica, Università di Pisa \& INFN, Sezione di Pisa, Largo Pontecorvo 3, I-56127 Pisa, Italy}

\author{Giovanni Villadoro}
\email{giovanni.villadoro@ictp.it}
\affiliation{INFN, Sezione di Trieste, Via Valerio 2, I-34127 Trieste, Italy}
\affiliation{Abdus Salam International Centre for Theoretical Physics, Strada Costiera 11, 34151, Trieste, Italy}

\date{\today}

\begin{abstract}
We present a proof-of-concept numerical study of the real-time topological rate at non-zero momentum in quenched lattice QCD at a temperature $T\simeq 1.24 \, T_c \simeq 360$ MeV, as an important step toward the determination of this quantity in full QCD. Our strategy, already applied to compute the sphaleron rate in pure Yang--Mills and in full QCD, extracts the rate from the resolution of an appropriate inverse problem, solved applying the Hansen--Lupo--Tantalo (HLT) method to the thermal Euclidean time-correlator of the topological charge density. This method requires to control three different limits: continuum limit, limit of vanishing smearing width used in the HLT inverse problem resolution, and limit of vanishing smoothing radius used in the topological charge density correlator computation. Our lattice calculation is based on the standard Wilson discretization for the gauge action, and on three gauge ensembles with up to $N_\tau=16$ temporal points to achieve a controlled continuum limit. In all cases we employed an aspect ratio $LT=4$, which allowed us to compute the topological rate up to momenta as large as $p/T \sim 10$.
\end{abstract}

\maketitle

\section{Introduction}

In Yang--Mills theories, the non-trivial topological features of the gauge group manifold have important theoretical and phenomenological consequences. An example of current interest in high-energy particle physics is provided by high-temperature real-time topological transitions in Quantum Chromo-Dynamics (QCD). Indeed, the so-called \emph{topological rate} $\Gamma_\top$, the mean square rate of thermal topology-changing transitions for unit time and spatial volume, plays a key role both in hadron physics and axion cosmology. This quantity is defined as:
%
\beq
\label{eq:gammatop_def_realtime}
\Gamma_\top(\omega,p)
= \int \dd t \, \dd^3 x \, \ee^{\ii x_\mu p^\mu}\braket{q(\vec{x},t)q(\vec{0},0)}_T \, .
\eeq
%
Here $p_\mu=(\omega,\vec{p}=p\,\hat{n})$ is the four momentum, $q(\vec{x},t)$ is the topological charge density
\bee\label{eq:topchargedens_cont}
q(\vec{x},t)=\frac{1}{32\pi^2}\varepsilon^{\mu\nu\rho\sigma}\Tr\left[G_{\mu\nu}(\vec{x},t) G_{\rho\sigma}(\vec{x},t)\right],
\eee
defined in terms of the gluon field strength $G_{\mu\nu} = \partial_\mu A_\nu - \partial_\nu A_\mu - \ii [A_\mu,A_\nu]$, with $A_\mu=A_\mu^a T_a$ and $2\Tr[T_a T_b]=\delta_{ab}$, while the notation $\braket{\dots}_T$ stands for the thermal average at a temperature $T$:
\bee
\braket{O}_T = \frac{\Tr\left[O\exp\{-\mathcal{H}_{\QCD}/T\}\right]}{\Tr\left[\exp\{-\mathcal{H}_{\QCD}/T\}\right]},
\eee
with $\mathcal{H}_\QCD$ the QCD Hamiltonian.

In the context of axion cosmology, the topological rate has been recently recognized as a fundamental input to compute the QCD axion thermal rate in the early Universe~\cite{Notari:2022ffe}. The latter quantity is necessary to estimate the relic axion abundance created by QCD thermal topological transitions from the resolution of the Boltzmann equation for the axion distribution function~\cite{Notari:2022ffe,Bianchini:2023ubu,OHare:2024nmr,Berghaus:2025dqi,Asadi:2025cvm,Bouzoud:2026rur,Arza:2026rsl,Barbieri:2026ewj}. In the context of hadron physics, instead, the $p_\mu \to 0$ limit of the topological rate, the so-called \emph{sphaleron rate},
\beq
\Gamma_\S = \lim_{\substack{\omega\,\to\,0 \\ p\,\to\,0}}  \Gamma_\top(\omega,p) = \int \dd t \, \dd^3 x \, \braket{q(\vec{x},t)q(\vec{0},0)}_T,
\eeq
is key to understand the \emph{Chiral Magnetic Effect} (CME)~\cite{Kharzeev:2007jp,Fukushima:2008xe, Kharzeev:2013ffa}. This phenomenon is expected to take place whenever thermal topological transitions occur in a hot and dense strongly-interacting quark-gluon medium in the presence of an external magnetic field, an extreme environment that is experimentally produced in heavy-ion colliders. Due to the chiral anomaly, a topological transition creates an imbalance of chiral fermion excitations, which align along the magnetic field and provide the source for an electric current flowing in the medium along the magnetic field direction (the CME current). Theoretical predictions for the CME current are of utmost importance to guide the intense experimental activities targeting CME detection~\cite{Kharzeev:2024zzm}. In this respect, first-principles QCD inputs for $\Gamma_\S$ are fundamental to compute the equilibration rate of chiral imbalances in the quark-gluon plasma~\cite{McLerran:1990de}. 

Despite its theoretical and phenomenological significance for strong interactions and axion physics, the determination of the topological rate in QCD from first principles remains a rather unexplored realm. For asymptotically high temperatures, the real-time topological rate can be investigated at low momenta with classical simulations of thermal effective theories, see, e.g.,~\cite{Moore:2010jd,DOnofrio:2014rug,Guin:2026kbp} and references therein, and at moderate and high momenta with perturbative methods, see, e.g.,~\cite{Notari:2022ffe,Bouzoud:2026rur} and references therein. However, the topological rate is inherently a non-perturbative quantity due to its topological nature. Non-perturbative effects in QCD can survive up to very high temperatures in certain observables~\cite{Bonati:2015vqz,Petreczky:2016vrs,Borsanyi:2016ksw,Trunin:2015yda,Burger:2017xkz,Bonati:2018blm,Burger:2018fvb,Lombardo:2020bvn,Athenodorou:2022aay,Kotov:2025ilm,Bresciani:2025vxw,Ce:2025ihd,Bresciani:2025mcu}, and are expected to be dominant at temperatures closer to the QCD chiral crossover, which is the relevant range for many applications to axion phenomenology. Therefore, it would be desirable to address the determination of the topological rate also from a fully non-perturbative first-principles method. In this respect, numerical Monte Carlo simulations of lattice QCD provide the ideal framework. However, they present significant challenges when it comes to the calculation of real-time quantities. Indeed, the lattice approach is based on the Euclidean formulation of QCD, where analytic continuation from real time $t$ to imaginary time $\tau = \ii t$ is used to express thermal averages in terms of Euclidean functional integral expectation values, amenable to be evaluated via importance sampling Monte Carlo methods once the continuum space-time is discretized. In the Euclidean formulation, $\Gamma_\top$ is inaccessible using its real-time definition in Eq.~\eqref{eq:gammatop_def_realtime}, and one has instead to rely on the Kubo formula~\cite{Meyer:2011gj,Meyer:2015wax,Lowdon:2022keu}:
\bee\label{eq:Kubo_gamma_top}
\Gamma_\top(\omega,p) = \frac{2}{1-\ee^{-\omega/T}} \rho(\omega,p).
\eee
This expression relates the topological rate to a Euclidean quantity, the \emph{spectral density} $\rho(\omega,p)$ of the thermal Euclidean time correlator of the topological charge density $G_{\E}(\tau,p)$:
\beq
\label{eq:def_spec_dens}
G_{\E}(\tau,p) &=& - \int_0^{\infty} \frac{\dd \omega}{\pi} \rho(\omega,p) K(\omega,\tau),\\
\nonumber\\
K(\omega,\tau) &=& \frac{\cosh\left(\frac{\omega}{2T}-\omega\tau\right)}{\sinh\left(\frac{\omega}{2T}\right)},\\
\nonumber\\
G_{\E}(\tau,p) &=& \int \dd^3 x \, \ee^{\ii \vec{p}\cdot \vec{x}} \, \braket{q(\vec{x},\tau) q(\vec{0},0)}_{\E}.
\eeq
Here, the notation $\braket{\dots}_{\E}$ simply stands for the Euclidean path integral expectation value computed in the presence of a compactified time direction with length $1/T$ (where, as usual, periodic/anti-periodic boundary conditions are taken for the gluon/quark fields):
\bee
\braket{O}_{\E} = \frac{ \int [\dd A \,\dd \overline{\psi} \,\dd \psi] \, O \, \exp\left\{-\int_0^{1/T} \dd \tau \int \dd^3 x \, \mathcal{L}_{\E}\right\}}{ \int [\dd A \,\dd \overline{\psi} \,\dd \psi] \, \exp\left\{-\int_0^{1/T} \dd \tau \int \dd^3 x \, \mathcal{L}_{\E}\right\}} \, ,
\eee
with $\mathcal{L}_{\E}$ the QCD Lagrangian in Euclidean space-time. Notice that Eq.~\eqref{eq:Kubo_gamma_top} reduces to the customary Kubo equation for the sphaleron rate in the $p_\mu\to 0$ limit:
\beq
\lim_{\substack{\omega\,\to\,0 \\ p\,\to\,0}} \Gamma_\top(\omega,p) = \Gamma_\S= 2T\lim_{\omega\,\to\,0}\frac{\rho(\omega,0)}{\omega}.
\eeq
The challenging aspect of using Eq.~\eqref{eq:Kubo_gamma_top} to determine the topological rate is that one needs $\rho(\omega,p)$, but what one is actually capable of computing on the lattice is $G_{\E}(\tau,p)$, i.e., the convolution of $\rho(\omega,p)$ with the known kernel function $K(\omega,\tau)$. Inverting Eq.~\eqref{eq:def_spec_dens} with controlled systematic errors is numerically very hard because $G_{\E}(\tau,p)$ is only known within statistical errors and for a discrete set of momenta and time separations due to the employed discretized space-time. Thus, with a naive inversion procedure, a tiny variation in the input data for $G_{\E}(\tau,p)$ would produce a huge variation in $\rho(\omega,p)$, and thus uncontrolled errors on $\Gamma_\top$. On general grounds, the extraction of $\rho(\omega,p)$ from the knowledge of $G_{\E}(\tau,p)$ falls under a broad class of mathematically ill-conditioned problems known as \emph{inverse problems}, which appear in several different physical contexts (see Ref.~\cite{Rothkopf:2022fyo} for a recent review focusing on lattice field theories). The inherent difficulty in solving inverse problems numerically is the main obstacle that has hindered progress in lattice studies of real-time topological transitions. Indeed, so far only the zero-momentum rate has been investigated on the lattice and, until a few years ago (see below), all non-perturbative studies were limited to the pure-gauge case~\cite{Kotov:2018aaa, Altenkort:2020axj, BarrosoMancha:2022mbj}.

Fortunately, in recent years several strategies have been proposed to solve inverse problems numerically on the lattice, such as strategies based on sum rules~\cite{Boito:2022njs}, on Bayesian approaches~\cite{Horak:2021syv,DelDebbio:2021whr,Candido:2023nnb}, on perturbative-motivated Ans\"{a}tze of the spectral density~\cite{Altenkort:2020axj,Altenkort:2020fgs,Altenkort:2022yhb,Altenkort:2023oms}, on the Tikhonov regularization~\cite{Tikhonov:1963aaa,Astrakhantsev:2018oue,Astrakhantsev:2019zkr}, on the model-independent Backus--Gilbert approach~\cite{BackusGilbert1968:aaa,Brandt:2015aqk,Brandt:2015sxa}, on integral transforms~\cite{Bruno:2024fqc,Tsuji:2026zku,Giusti:2026mcy}, on a bootstrap-like approach~\cite{Abbott:2026wdw}, or on a machine-learning-based strategy~\cite{DeSantis:2026vqg}. In this context, a recent novel approach that has revealed to be particularly successful is the Hansen--Lupo--Tantalo (HLT) method~\cite{Hansen:2019idp}, a non-trivial modification of the Backus--Gilbert method that has been employed quite extensively and for several different physical applications~\cite{Bulava:2021fre,ExtendedTwistedMassCollaborationETMC:2022sta,Frezzotti:2023nun,Evangelista:2023fmt,ExtendedTwistedMass:2024myu,Frezzotti:2024kqk,Bennett:2024cqv,Almirante:2024lqn,DeSantis:2025yfm,DeSantis:2025qbb,TELOS:2025ash,DiCarlo:2025mnm,DiPalma:2026fni,Lupo:2026vdj}.\footnote{The relation of HLT (and Backus--Gilbert methods in general) with Bayesian approaches to inverse problem resolution is discussed in~\cite{DelDebbio:2024lwm}.} In Ref.~\cite{Bonanno:2023ljc}, a new strategy to compute the sphaleron rate from the lattice  has been introduced, based on the inversion of Eq.~\eqref{eq:def_spec_dens} at $p=\omega=0$ via HLT. In that study, it was shown that the new method, applied to the pure Yang--Mills theory, yielded agreeing results with the previous pure-gauge study~\cite{BarrosoMancha:2022mbj}, where the authors evaded the resolution of the inverse problem by adopting a tailored technique to compute $\Gamma_\S$. Shortly after, the new method~\cite{Bonanno:2023ljc} was applied to full QCD, allowing the first non-perturbative determination of the sphaleron rate with dynamical quarks~\cite{Bonanno:2023thi}.

The present study aims at further pushing forward the program started by Refs.~\cite{Bonanno:2023ljc,Bonanno:2023thi}, extending the lattice investigation of real-time topological transitions to the non-zero momentum case. This is of particular importance for axion phenomenology, where the momentum-dependent topological rate determines the momentum-dependent axion rate entering the Boltzmann equation for the axion number distribution function. As pointed out in~\cite{Notari:2022ffe}, taking into account the momentum-dependence of the axion rate is key for a solid estimation of the hot axion relic abundance: axion production could either never reach equilibrium, or it could reach equilibrium at different times depending on the axion momentum (as higher momenta decouple later from the cosmological medium than lower ones). These arguments thus explain the necessity of computing the axion number from a momentum-dependent Boltzmann equation, for which the momentum-dependent topological rate is a key input.

The goal of this paper is to explore the feasibility of the method used in~\cite{Bonanno:2023ljc,Bonanno:2023thi} to compute the topological rate in a simplified setup, in view of a future application to the much more computationally demanding case of full QCD. For this reason, we will focus on just one value of the temperature, and on the quenched theory without dynamical fermions. This manuscript is organized as follows: in Sec.~\ref{sec:setup} we describe our numerical setup; in Sec.~\ref{sec:res} we present our lattice results for the topological rate as a function of the four momentum; finally, in Sec.~\ref{sec:conclu} we draw our conclusions, and discuss future perspectives for the extension of this work to full QCD. 

\section{Numerical setup}\label{sec:setup}

\subsection{Lattice action and simulation parameters}\label{sec:discretization}

We discretize the continuum space-time on an hyper-cubic lattice with lattice spacing $a$, $N_s$ spatial points and $N_\tau$ temporal points. The spatial and temporal sizes in physical units are denoted with $L=a N_s$ and $L_0=a N_\tau=1/T$ respectively. We adopt the customary Wilson plaquette action to define our lattice Yang--Mills theory,
\beq\label{eq:wilson_action}
\mathcal{S}_{\scriptscriptstyle{\rm YM}}[U] = -\frac{\beta}{3} \sum_{x,\mu>\nu}\mathrm{Re} \Tr \left[ U_{\mu\nu}(x) \right],
\eeq
with $U_{\mu\nu}(x)=U_\mu(x) U_\nu(x+a\hat{\mu})U^\dagger_\mu(x+a\hat{\nu})U^\dagger_\nu(x)$ the plaquette, $U_\mu(x) \in \mathrm{SU}(3)$ the gauge links, and $\beta=6/g^2$ the inverse bare gauge coupling.

Simulation parameters are inherited from the previous study~\cite{Bonanno:2023ljc}, with the only difference that we increase the spatial volume. In particular, our three simulation points lie on a \emph{Line of Constant Physics} (LCP) with fixed $L$ and $T$, where the spatial size was fixed adopting an aspect ratio $LT=N_s/N_\tau=4$ (as opposed to 3 in~\cite{Bonanno:2023ljc}), while the temperature was fixed to $T\simeq 1.24 \,T_c$, with $T_c$ the SU(3) critical deconfinement temperature. In finite-temperature gauge theories, lattice artifacts are controlled by $aT=1/N_{\tau}$, thus $N_{\tau} \gtrsim\mathcal{O}(10)$ is typically considered a safe threshold to avoid large lattice artifacts. For this reason, we chose $N_\tau=12,14,16$. The absolute scale was set using the determinations of the Sommer parameter $r_0$ of Ref.~\cite{Necco:2001xg}.\footnote{In Ref.~\cite{Bonanno:2023ljc} it was checked that the different parameterization of $a(\beta)/r_0$ of Ref.~\cite{Francis:2015lha} gave perfectly compatible results for the LCP within errors.} Finally, the functional integral was sampled using a combination of heat-bath~\cite{Creutz:1980zw,Kennedy:1985nu} and over-relaxation~\cite{Creutz:1987xi} updating algorithms implemented \emph{\`a l\`a} Cabibbo--Marinari~\cite{Cabibbo:1982zn}, i.e., updating the three diagonal SU(2) subgroups of SU(3). A single updating step consists in one lattice sweep of heat-bath and four lattice sweeps of over-relaxation. The total accumulated statistics and the simulation parameters are reported in Tab.~\ref{tab:simulation_summary}.

\begin{table}[!t]
\begin{center}
\begin{tabular}{|c|c|c|c|c|c|c|}
\hline
$N_s$ & $N_\tau$ & $\beta$ & $a(\beta)/r_0$ & $L/r_0$ & $r_0T$ & Statistics\\
\hline
48 & 12 & 6.440 & 0.09742(97) & 1.141(11) & 3.507(35) & 53.9k \\
56 & 14 & 6.559 & 0.08364(84) & 1.139(11) & 3.513(35) & 21.7k \\
64 & 16 & 6.665 & 0.07309(73) & 1.140(11) & 3.508(35) & 15.7k \\
\hline
\end{tabular}
\end{center}
\begin{center}
\begin{tabular}{|c|c|c|c|c|c|}
\hline
$N_\tau$ & $\beta$ & $a(\beta)$~[fm] & $L$~[fm] & $T$~[MeV] & $T/T_c$ \\
\hline
12 & 6.440 & 0.04598(67) & 2.207(32) & 357.6(5.2) & 1.244(18) \\
14 & 6.559 & 0.03948(58) & 2.211(32) & 357.0(5.2) & 1.242(18) \\
16 & 6.665 & 0.03450(50) & 2.208(32) & 357.5(5.2) & 1.244(18) \\
\hline
\end{tabular}
\end{center}
\caption{Collection of simulation parameters. The lattice spacing in units of the Sommer scale $r_0$ was obtained from an interpolation of the determinations of $r_0/a(\beta)$ of Ref.~\cite{Necco:2001xg} in the range $5.7 \le \beta \le 6.92$, cf.~Eq.~(2.6) in that paper. To express $a$ and $L$ in fm, and the temperature in MeV and in units of $T_c$, we used the values of $r_0$ and $w_0$ in fm of, respectively, Refs.~\cite{Sommer:2014mea} and~\cite{Borsanyi:2012zs}, and the result for $w_0T_c$ of~\cite{Borsanyi:2022xml}. The total collected statistics is expressed in thousands (k). Measures were collected every 20 updating steps after skipping 2000 thermalization steps (for the definition of a single updating step see the main text of Sec.~\ref{sec:discretization}).}
\label{tab:simulation_summary}
\end{table}

\subsection{Topological charge density correlator and the role of smoothing}\label{subsec:smoothing_role}

In this study, we adopt the clover discretization of the continuum topological charge density~\eqref{eq:topchargedens_cont}, a gluonic lattice formulation possessing definite parity:
\bee\label{eq:clover_charge_def}
q_{\LL}(x) = \frac{1}{32\pi^2} \sum_x \sum_{\mu\nu\rho\sigma}\varepsilon_{\mu\nu\rho\sigma}\Tr\left[C_{\mu\nu}(x)C_{\rho\sigma}(x)\right],
\eee
with
\bee
\begin{aligned}
C_{\mu\nu}(x) = \frac{1}{4} &\Im\left\{ U_{\mu\nu}(x) + U_{-\nu, \mu}(x)\right.\\
&\,\,\left. + U_{\nu, -\mu}(x) + U_{-\mu, -\nu}(x)\right\}
\end{aligned}
\eee
the four-leaf clover, and where plaquettes with negative indices are built according to $U_{-\mu}(x) = U^\dagger_{\mu}(x-a\hat{\mu})$.

Then, the correlation function $G_{\E}(\tau,p)$ is built in terms of the time profiles of the topological charge, defined as:
\beq
Q_{\LL}(\tau,p) = \sum_{\vec{x}} \ee^{\ii \vec{p}\cdot\vec{x}} q_{\LL}(\vec{x}, \tau).
\eeq
More precisely, we computed:
\beq\label{eq:corr_lat_def}
a^5 G_{\E}(\tau,p) = \frac{1}{N_s^3}\braket{Q_{\LL}(\tau_1,p)Q_{\LL}^*(\tau_2,p)}_{\E},
\eeq
where we averaged $G_{\E}$ over all couples $(\tau_1, \tau_2)$, $\tau_2\ge \tau_1$, that corresponded to the same time separation
\bee
\tau=
\begin{cases}
\tau_2-\tau_1, &\qquad\mathrm{if}\quad \tau_2 - \tau_1 \le \dfrac{1}{2T},\\
\\
\dfrac{1}{T}- (\tau_2-\tau_1), &\qquad\mathrm{if}\quad \tau_2 - \tau_1 >\dfrac{1}{2T}.
\end{cases}
\eee
Due to the periodicity in the time direction, one can limit to take $\tau \in [0,1/(2T)]$, since $G_{\E}(\tau)=G_{\E}(1/T-\tau)$. Concerning the spatial momentum, for this first feasibility study we simply limited to take it along the x-axis, $\vec{p}=(p,0,0)$, and leave the investigation of other possible choices for future work. Let us also recall that the modulus $p$ in lattice units is discretized in units of the inverse spatial lattice size,
\beq\label{eq:discr_mom}
ap = \frac{2 \pi}{N_s} k,  \qquad k=0,1,2,\dots,N_s/2 \, .
\eeq

The charge time profiles entering Eq.~\eqref{eq:corr_lat_def} are not computed directly from the gauge field configurations sampled by the Monte Carlo algorithm, but are instead evaluated after applying a \emph{smoothing} procedure. This is necessary to get rid of the multiplicative renormalization factor that would affect the lattice topological charge density when computed on unsmoothened fields~\cite{Campostrini:1988cy,Campostrini:1989dh,DElia:2003zne,Vicari:2008jw}:
\bee
Q_{\LL} = \sum_{x} q_{\LL}(x) = Z(\beta) Q, \quad \,\,\, Q=\int \dd^4 x \, q(x) \in \mathbb{Z},
\eee
with $Q$ the integer-valued continuum topological charge. However, if on one hand smoothing has such beneficial effect, on the other hand it introduces an unphysical new scale, the \emph{smoothing radius} $\Rs$. This represents the scale below which ultra-violet fluctuations are smoothened away, and unavoidably alters the short-distance behavior of $G_{\E}(\tau,p)$. In particular, it is well-known that, in the continuum theory, $G_{\E}(\tau,p)< 0$ for every $\tau > 0$ due to \emph{reflection positivity} and the parity-odd nature of the topological charge density~\cite{Alles:1997ae,Vicari:1999xx,Horvath:2005cv,Vicari:2008jw,Chowdhury:2012sq,Fukaya:2015ara,Mazur:2020hvt}. On the lattice, this condition fails when using smeared sources in the correlator, leading to a positive $G_{\E}(\tau,p)$ as soon as the time separation $\tau$ is smaller than $\Rs$. To recover the proper physical correlation function at all time separations, one needs first to take the continuum limit $a\to 0$ at fixed $\Rs$, and then to take the $\Rs\to 0$ limit. Taking these two limits restores the full negativity of $G_{\E}(\tau,p)$ at all positive time separations.

In the literature, several smoothing algorithms have been employed to study the topological properties of Yang--Mills theories, such as cooling~\cite{Berg:1981nw,Iwasaki:1983bv,Itoh:1984pr,Teper:1985rb,Ilgenfritz:1985dz,Campostrini:1989dh,Alles:2000sc}, stout smearing~\cite{APE:1987ehd, Morningstar:2003gk} or gradient flow~\cite{Luscher:2010iy}. Once the smoothing radii are matched to one another, all these methods have been shown to give consistent results~\cite{Alles:2000sc, Bonati:2014tqa, Alexandrou:2015yba}. In this work we have adopted Wilson cooling for its simplicity and numerical cheapness. This is an iterative smoothing algorithm designed to drive a given gauge field configuration closer to a local minimum of the Wilson plaquette action~\eqref{eq:wilson_action}. In practice, a single cooling step is achieved by iteratively aligning each link $U_{\mu}(x)$ to its local staple. With cooling, as with any other smoothing method, $\Rs$ is proportional to the square root of the number of smoothing steps, reflecting the diffusive nature of the procedure. In particular, the smoothing radii corresponding to $\ncool$ steps of Wilson cooling and to the Wilson gradient-flow time $t_{\sst{\rm GF}}$ are
\beq\label{eq:cooling_gradientflow_matching}
\Rs = a \sqrt{\frac{8}{3}\ncool} = \sqrt{8t_{\sst{\rm GF}}},
\eeq
where the second equality provides a matching between the two smoothing procedures~\cite{Bonati:2014tqa} (see also~\cite{Alexandrou:2015yba,Alexandrou:2017hqw}). The equivalence of cooling and gradient flow for the purpose of computing $G_{\E}(\tau,p=0)$ (after the double extrapolation $a\to 0$ followed by $\Rs \to 0$) has been explicitly shown in~\cite{Bonanno:2023ljc}.

There is, however, an important point to be addressed concerning the use of cooling. The imaginary part of $G_{\E}(\tau,p)$ is not identically zero for $p\neq 0$ on individual gauge configurations, but its average must vanish within statistical uncertainty because of the time-reversal symmetry of the Wilson action. However, some implementations of cooling explicitly break time-reversal symmetry, thus $G_{\E}(\tau,p)$ can acquire a non-vanishing imaginary part at finite smoothing radius. In particular, this is the case for the implementation adopted in this work, for which we find violations up to several standard deviations of $\mathrm{Im}[G_{\E}(\tau,p)]=0$, although the imaginary part remains much smaller than $\mathrm{Re}[G_{\E}(\tau,p)]$ in most cases (typically, one order of magnitude smaller). Time-reversal symmetry is broken because, for computational efficiency, gauge links are cooled following an even-odd ordering. More details are given in Appendix~\ref{app:cooling}, in which we also compare this implementation with gradient flow and with a symmetry-preserving version of cooling. We explicitly verify that, after matching $R_s$, all three smoothing procedures give perfectly compatible results for the real part of $G_{\E}(\tau,p)$ at our coarsest lattice spacing. In light of these results, we employ the symmetry-breaking implementation of cooling in the rest of this work because it is more computationally efficient than the other cooling implementation.

\subsection{Topological rate from HLT method}

This section is devoted to explain how we applied the HLT method to invert Eq.~\eqref{eq:def_spec_dens} and obtain the topological rate. Let us start by introducing the following notation for the topological rate:
\beq
\Gamma_\top(\omega,p) &=& \frac{\omega/T}{1-\ee^{-\omega/T}} \, 2T \left[\frac{\rho(\omega,p)}{\omega}\right] \\
\nonumber&&\\
\label{eq:true_rate}
&\equiv& f(\omega) \, 2T\left[\frac{\rho(\omega,p)}{\omega}\right],
\eeq
with $f(\omega=0)=1$. Let us also introduce the modified kernel $K^\prime(\omega,\tau)$ as:
\beq
G_{\E}(\tau,p) &=& - \int_0^{\infty} \frac{\dd \omega}{\pi} \left[\frac{\rho(\omega,p)}{\omega}\right] \omega K(\omega,\tau)\\
\nonumber&&\\
\label{eq:true_inversion}
&\equiv& - \int_0^{\infty} \frac{\dd \omega}{\pi} \left[\frac{\rho(\omega,p)}{\omega}\right] K^\prime(\omega,\tau).
\eeq
These steps are useful to express everything in terms of a quantity, $\rho(\omega,p)/\omega$, which is finite for every $\omega$, including $\omega \to 0$. At the same time, since $K(\omega,\tau)\sim 1/\omega$ for $\omega\to 0$, the new kernel $K^\prime(\omega,\tau)$ is now non-singular in the origin. The plan now is to apply the HLT method to obtain $\rho(\omega,p)/\omega$ from the inversion of Eq.~\eqref{eq:true_inversion}, and then use Eq.~\eqref{eq:true_rate} to obtain $\Gamma_\top(\omega,p)$.

The ansatz solution for the spectral density is:
\beq\label{eq:HLT_ansatz}
\frac{\rhobar(\omegastar,p)}{\omegastar} = -\pi \sum_\tau g_\tau(\omegastar) G_{\E}(\tau,p),
\eeq
where the sum over $\tau$ means that we are summing over all available discrete time separations $\tau/a=n$ with $n=1,2,\dots,N_\tau/2$, and where the $g_\tau(\omegastar)$ are unknown coefficients to be determined via HLT. Inserting Eq.~\eqref{eq:true_inversion} in Eq.~\eqref{eq:HLT_ansatz}, one obtains the following consistency relation:
\beq\label{eq:HLT_smear_relation}
\frac{\rhobar(\omegastar,p)}{\omegastar} = \int_0^{\infty} \dd \omega  \, \Delta(\omega,\omegastar) \frac{\rho(\omega,p)}{\omega},
\eeq
with 
\beq
\Delta(\omega,\omegastar) = \sum_\tau g_\tau(\omegastar)K^\prime(\omega,\tau).
\eeq
The consistency relation \eqref{eq:HLT_smear_relation} has the form of a \emph{smearing relation}, where $\Delta(\omega,\omegastar)$ acts as the smearing kernel. Thus, our ansatz solution is nothing but a \emph{smeared} solution, related to the true solution via Eq.~\eqref{eq:HLT_smear_relation}. The idea is now to look for the ``optimal'' (in what sense will be clarified in a moment) set of coefficients $g_\tau(\omegastar)$ that will make $\Delta(\omega,\omegastar)$ as much peaked as possible around $\omegastar$, so that $\frac{1}{\omegastar}\rhobar(\omegastar,p)$ is the optimal approximation of the true solution $\frac{1}{\omega}\rho(\omega,p)\big\vert_{\omega\,=\,\omegastar}$. In the limit in which $\Delta(\omega,\omegastar)=\delta(\omega-\omegastar)$, one would have an exact equality between the ansatz and the true solution, but this is of course never reachable in a practical calculation. What one can do is instead to retain a finite width in the smearing kernel, and try to make it as small as possible to render its impact negligible within errors. The HLT method exactly provides a numerical strategy to accomplish this task, as it allows to determine the $g_\tau(\omegastar)$ coefficients for a given input smearing width $\sigma$, whose value can be changed to check the impact of smearing on the topological rate.

Let us now describe in details how the optimal $g_{\tau}(\omegastar)$ coefficients are determined, and in what sense they are ``optimal''. The main idea behind the HLT method is to obtain them from the resolution of a minimization problem with the following functional:
\bee
F_\lambda[g] = (1-\lambda) A_\alpha[g] +  \lambda B[g], \,\quad  \lambda\in [0,1).
\eee
Here, the functional $A_\alpha[g]$ quantifies the distance of the actual smearing kernel $\Delta(\omega,\omegastar)$ from a chosen target kernel, $\delta_\sigma(\omega,\omegastar)$, possessing a characteristic width $\sigma$ around the desired energy $\omegastar$:
\beq
A_\alpha[g] = \int_0^{\infty} \dd\omega \, \left\vert \Delta(\omega,\omegastar) - \delta_\sigma(\omega,\omegastar)\right\vert^2 \, \ee^{a \alpha \omega},
\eeq
with $0 \le\alpha < 2$ (we chose $\alpha=2^-=1.99$ throughout this study). Instead, the functional $B[g]$ takes into account the statistical errors on the lattice correlation function:
\beq
B[g] = \frac{1}{G_{\E}^2(0,p)} \sum_{\tau,\tau^\prime}\mathrm{Cov}(\tau,\tau^\prime; p) g_{\tau}(\omegastar) g_{\tau^\prime}(\omegastar),
\eeq
with $\mathrm{Cov}(\tau,\tau^\prime;p)$ denoting the covariance matrix of $G_{\E}(\tau,p)$, and where the factor of $1/G^2_{\E}(0,p)$ is just a normalization. The relative weight of the two terms is regulated by the free parameter $\lambda$. When $\lambda\to 0$, $A_\alpha$ dominates $F_\lambda$, and one is effectively enforcing the constraint that $\Delta$ is actually close to $\delta_\sigma$, but the $g_\tau(\omegastar)$ will oscillate uncontrollably, and thus the resulting estimate of the spectral density will be affected by huge statistical uncertainties. As $\lambda$ is increased, the contribution of the $B$ term starts to become important, and this has the effect of damping the fluctuations affecting $g_\tau(\omegastar)$, making the statistical uncertainties on $\rhobar(\omegastar,p)/\omegastar$ sensible. However, when $\lambda\to 1$, the weight of $B$ will overcome the weight of $A_\alpha$ in $F_\lambda$, thus yielding an unconstrained smearing kernel $\Delta$, whose shape and width around the desired energy $\omegastar$ are not under control. This will eventually lead to a bias in the obtained values of the topological rate due to spurious contributions coming from the true spectral density at energies $\omega \ne \omegastar$. The idea is to look for an optimal balance between these two regimes, and in particular to stay within a statistically dominated regime where systematic effects are negligible within errors, but where at the same time statistical fluctuations are under control. The optimal trade off between statistical and systematic errors is found from the so-called \emph{stability analysis}. One minimizes the functional $F_\lambda[g]$ for several values of $\lambda$, and then looks for a plateau of $\Gamma_\top$ as a function of the figure of merit:
\beq
d_\lambda \equiv \sqrt{\frac{A_0[g_\lambda^\star]}{A_0[0]}},
\eeq
with $g_\lambda^\star$ denoting the result of the minimization of $F_\lambda[g]$. The quantity $d_\lambda$ expresses the normalized distance between the actual smearing kernel $\Delta$ and the target smearing kernel $\delta_\sigma$ for a given value of $\lambda$. The statistically-dominated regime is characterized by small values of $d_\lambda$, as the reconstructed smearing kernel will be reasonably close to the target one, and thus it will possess a characteristic width of the order of $\sigma$ around $\omegastar$. In all cases, to quote our final value for $\Gamma_\top$, we followed the same procedure illustrated in Ref.~\cite{ExtendedTwistedMassCollaborationETMC:2022sta}, cf.~Eqs.~(37) and~(38) there. First, one selects two values $\lambda_2<\lambda_1$ within the statistically-dominated regime. Then, the central value and statistical error on $\Gamma_\top$ are obtained from $\lambda_1$. Finally, $\lambda_2$ is used to estimate a systematic error, which is then summed in quadrature to the statistical one. The systematic error is computed as follows:
\beq
\Delta_{\sst{\rm syst}} &=& \bigg\vert \, \Gamma_\top\big\vert_{\lambda_1} - \Gamma_\top\big\vert_{\lambda_2} \, \bigg\vert \times \mathrm{erf}(D/\sqrt{2}), \\
\nonumber\\
D &=& \frac{\bigg\vert \, \Gamma_\top\big\vert_{\lambda_1} - \Gamma_\top\big\vert_{\lambda_2} \, \bigg\vert}{\sqrt{\Delta^2_{\sst{\rm stat}}\big\vert_{\lambda_1}+\Delta^2_{\sst{\rm stat}}\big\vert_{\lambda_2}}},
\eeq
with $\mathrm{erf}$ the standard error function. Statistical errors on $\Gamma_\top$ are instead obtained from a standard binned bootstrap procedure.

Finally, let us comment on the target smearing kernel, and on the dependence of the final result on its width (which we recall is an input of the HLT method). Generally speaking, the target smearing kernel can be selected with some degree of arbitrariness (in some cases, different choices can be mapped into one another, see~\cite{Jay:2026qoh}). Here we adopt the same choice of target kernel of Refs.~\cite{Bonanno:2023ljc,Bonanno:2023thi}, which was inspired by the functional form of $K^\prime(\omega,\tau=\frac{1}{2T})$:
\beq\label{eq:target_kernel_def}
\delta_\sigma(\omega,\omegastar) = \frac{4}{\sigma \pi^2} \frac{x}{\sinh(x)}, \quad x = \frac{\omega-\omegastar}{\sigma}.
\eeq
This is a pseudo-Gaussian envelop with a characteristic width $\sigma$ around $\omega=\omegastar$. When the reconstruction of the smearing kernel via HLT has a small $d_\lambda$, also the smearing kernel $\Delta(\omega,\omegastar)$ will show a characteristic width of the order of $\sigma$ around the desired energy. By repeating the inverse problem resolution for various values of $\sigma$, one can study its impact on the final result for the topological rate. For smearing kernels that are symmetric around $\omegastar$ like~\eqref{eq:target_kernel_def}, one expects~\cite{Bulava:2021fre}:
\beq
\Gamma_\top\big\vert_\sigma = \Gamma_\top + C \sigma^2 + \mathcal{O}(\sigma^4),
\eeq
with $\Gamma_\top$ the true topological rate, and $\Gamma_\top\big\vert_\sigma$ the rate obtained for a finite smearing width. One can use this general prediction to perform a $\sigma \to 0 $ extrapolation of the numerical data for the rate.

\section{Numerical results}\label{sec:res}

In this section, we will present our results for the topological rate as a function of the energy $\omega$ and of the spatial momentum $p$. On general grounds, these two parameters can be varied independently, as the energy is decided by the point $\omegastar$ at which the HLT inversion is performed, while the momentum is decided by the one entering the correlation function that is being inverted. For applications to axion physics, however, one is interested in:
\bee
\omega = E_p = \sqrt{m_a^2 + p^2} \simeq p \qquad \text{(on-shell axion)},
\eee
as $\omega$ and $p$ exactly represent the energy and spatial momentum of the axion (we are here assuming that the axion mass $m_a$ is so small compared to $p$ that it can be safely neglected, see~\cite{GrillidiCortona:2015jxo,Notari:2022ffe}). In the following section, we will show both \emph{off-shell} results obtained varying $\omega$ and $p$ independently, as well as \emph{on-shell} results satisfying $\omega=p$, which are those of more direct phenomenological interest for the axion. As we shall see, the availability of off-shell results will be important to understand the observed behavior with $p$ of the on-shell topological rate.

As we have discussed in Sec.~\ref{sec:setup}, one has in principle to take three different limits: continuum limit $a\to 0$, vanishing smoothing radius limit $\Rs \to 0$ (to be taken after the continuum limit), and vanishing smearing width limit $\sigma \to 0$. Following the lines of Ref.~\cite{Bonanno:2023ljc}, we will follow two different strategies to crosscheck our determinations of $\Gamma_\top$. Both routes have been shown in Ref.~\cite{Bonanno:2023ljc} to give compatible results.

\noindent\textbf{Method 1.---}First, we will follow a strategy inspired by Refs.~\cite{Altenkort:2020axj,Kotov:2018aaa}, namely, we will first perform a double extrapolation ($a\to 0$ followed by $\Rs \to 0$) of the correlation functions $G_{\E}(\tau,p)$. Then, we will apply the HLT method to the double-extrapolated correlators to obtain $\Gamma_\top$ for several values of $\sigma$. This route presents two main downsides. One is that the correlators at finite lattice spacings will be defined for different values of the time separation in physical units $\tau T$. Since the double limit must be taken at fixed $\tau T$, this requires to interpolate the correlators obtained on the coarser lattices to the values of $\tau T$ of the finest one. The second one, as we shall see later, is that the range of available smoothing radii to perform the $\Rs \to 0$ limit shrinks as one reduces $\tau$, thus in practice the double-limit can only be computed down to a certain minimum value of $\tau$.

\noindent\textbf{Method 2.---}The second strategy we will follow, introduced in Ref.~\cite{Bonanno:2023ljc}, instead consists in performing the HLT inversion directly on the finite-lattice-spacing and finite-$\Rs$ correlators, and perform all the three limits ($a \to 0$, $\Rs \to 0$ and $\sigma \to 0$ directly on the topological rate. This procedure avoids the two downsides of the first approach earlier discussed, and it was found in Ref.~\cite{Bonanno:2023ljc} to give agreeing results with Method 1.

\subsection{Topological susceptibility and choice of smoothing radius range}\label{sec:rate_calc_1}

As earlier explained, we will need to perform a zero-smoothing-radius extrapolation $\Rs \to 0$ to obtain our final results for the topological rate. Following the same line of reasoning of Ref.~\cite{Bonanno:2023ljc}, we established the range of sensible smoothing radii as follows. For the minimum smoothing radius, we required that $\Rs$ was at least 2 lattice spacings for all simulation points. This requires to take $\ncool \ge 2$ as $\Rs(\ncool=2)=a\sqrt{16/3}\approx 2.3a$. Given that our coarsest simulation point has $N_\tau=12$, the latter choice thus corresponds to a minimum physical smoothing radius of $\Rs T = \sqrt{(16/3)} \times 1/12 \simeq 0.192$. The maximum smoothing radius was instead chosen so that $\Rs$ equals the maximum temporal distance one can have in the presence of a periodic temporal size with length $1/T$, i.e., $\Rs T = 1/2=0.5$. This in practice means that the maximum number of cooling steps was chosen to be $\ncool=13,18,24$ for $N_\tau=12,14,16$.

\begin{figure}[!t]
\centering
\includegraphics[scale=0.47]{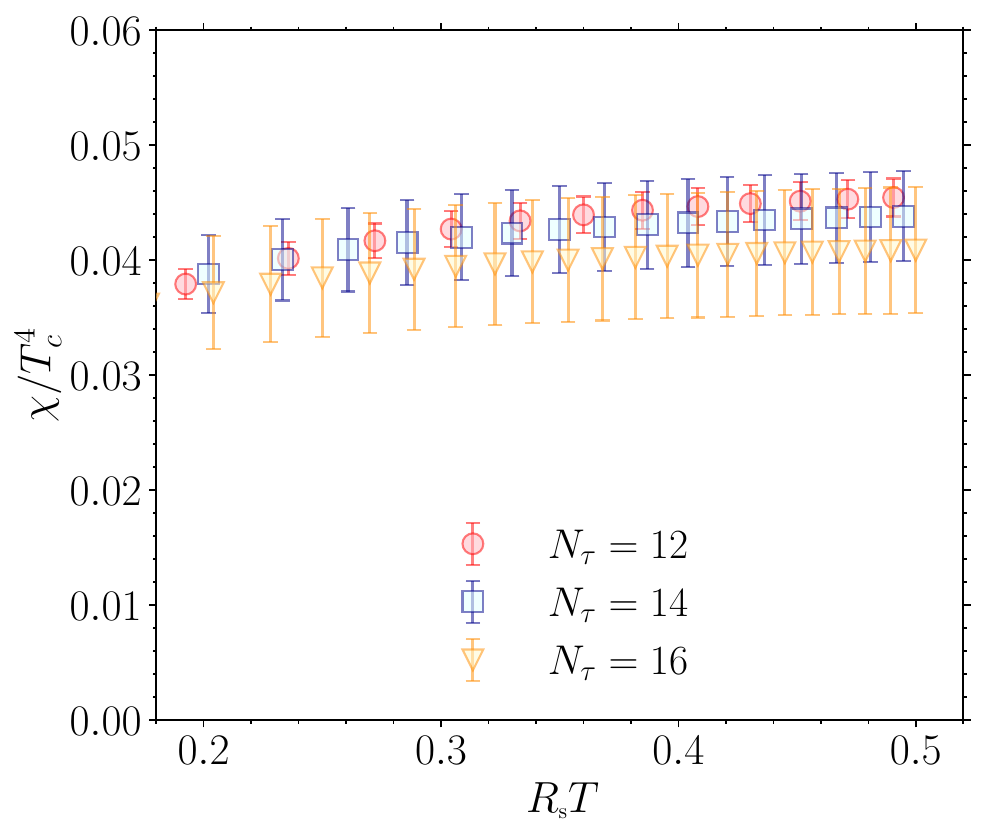}
\caption{Dependence of the finite-lattice-spacing determinations of the topological susceptibility $\chi/T_c^4$ on the smoothing radius $\Rs T$.}
\label{fig:chi_Rs_dep}
\end{figure}

In order to check that this choice of smoothing radii is large enough to correctly identify the correct topological background of all gauge configurations and, at the same time, not excessively large to destroy important topological fluctuations, we computed the continuum limit of the topological susceptibility:
\beq
\chi = \frac{T}{L^3}\braket{Q^2},
\eeq
with $Q$ the clover lattice topological charge in Eq.~\eqref{eq:clover_charge_def} computed after cooling, at fixed smoothing radius in physical units for several values of $\Rs T$. Results obtained for different $N_\tau$ and $\Rs T$ are interpolated using a cubic spline to achieve the same physical smoothing radius up to 4 digits. The dependence of $\chi/T_c^4$ on $\Rs T$ for all available lattice spacings is illustrated in Fig.~\ref{fig:chi_Rs_dep}.

\begin{figure}[!t]
\hspace*{-1.2\baselineskip}
\includegraphics[scale=0.54]{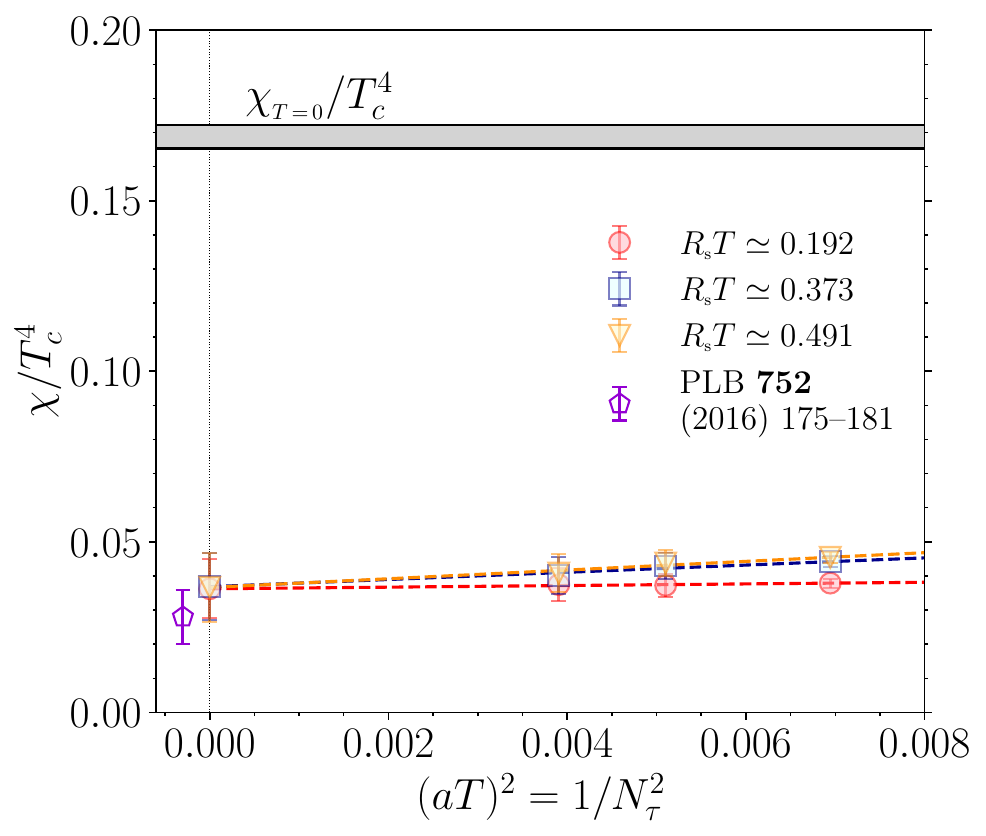}
\caption{Extrapolation toward the continuum limit of the topological susceptibility $\chi/T_c^4$ assuming linear corrections in $(aT)^2$, for three values of the smoothing radius $\Rs T$: the minimum and the maximum considered, and one exactly in between. The continuum limit is independent of $\Rs T$, as expected, and agrees with the continuum extrapolation of Ref.~\cite{Borsanyi:2015cka} obtained at the same temperature from $N_\tau \le 8$. For comparison, we also show the zero-temperature result $\chi_{_{\sst{T\,=\,0}}}/T_c^4$, computed taking $\chi_{_{\sst{T\,=\,0}}}/\sigma^2$ from Ref.~\cite{Bonanno:2025eeb} and $T_c/\sqrt{\sigma}$ from Ref.~\cite{Lucini:2012wq}.}
\label{fig:contlim_chi}
\end{figure}

In principle, if the smoothing range is sufficiently large to avoid UV contamination in the topological charge and, at the same time, not excessively large to destroy physical infrared fluctuations, one should see a smooth continuum limit of $\chi$ independent of the choice of $\Rs T$. At finite lattice spacing, instead, one could see a weak dependence of $\Rs T$ which should become less and less visible as $a \to 0$ (i.e., $\Rs T$ just controls the magnitude of lattice artifacts). The same procedure was also employed in the zero-temperature case to compute $\chi$ in $\mathrm{SU}(N)$ Yang--Mills theories~\cite{Ce:2015qha,Ce:2016awn,Bonanno:2025eeb}, and also in~\cite{Giusti:2018cmp} to compute $\chi$ at finite $T$ in the pure $\mathrm{SU}(3)$ gauge theory using the Wilson gradient flow as smoothing method, and performing Master Field simulations. The result is displayed in Fig.~\ref{fig:contlim_chi}. As it can be seen, in all cases we obtain a smooth continuum limit which turns out, as expected, to be independent of $\Rs$. Our final result reads $\chi/T_c^4=0.0363(87)$. This is also perfectly compatible with the continuum extrapolation obtained in Ref.~\cite{Borsanyi:2015cka} at the same temperature, $\chi/T_c^4 \simeq 0.028(8)$. The result of~\cite{Borsanyi:2015cka} was obtained from a continuum extrapolation of a gradient-flow-based determination of the susceptibility at fixed smoothing radius $\Rs = \sqrt{8t_{\scriptscriptstyle{\rm GF}}}=\frac{1}{\sqrt{2}} \frac{1}{T_c} \simeq 0.877 \frac{1}{T}$. The authors employed smaller temporal extents $N_\tau \le 8$, but also a tree-level Symanzik-improved gauge action. The remarkable agreement among our and Ref.~\cite{Borsanyi:2015cka}'s determinations is thus non-trivial. We therefore conclude that our choice $\Rs T \in [0.192,0.5]$ of smoothing radii is perfectly safe, and we will use it for the purpose of extrapolating the topological rate to $\Rs \to 0$. \emph{En passant}, we observe that our result for $\chi/T_c^4$ at $T/T_c\simeq 1.24$ is much smaller (about a factor of 6) than the zero-temperature susceptibility $\chi_{_{\sst{T\,=\,0}}}/T_c^4 \simeq 0.1687(34)$, computed using continuum results of Refs.~\cite{Lucini:2012wq,Bonanno:2025eeb} for, respectively, $\chi_{_{\sst{T\,=\,0}}}/\sigma^2$ and $T_c/\sqrt{\sigma}$, with $\sigma$ the string tension. This is in accord with general expectations based on the Dilute Instanton Gas Approximation, predicting $\chi$ to be rapidly suppressed in the high-temperature phase.

\subsection{Topological rate from the double-extrapolated correlator (Method 1)}

\begin{figure}[!t]
\centering
\includegraphics[scale=0.48]{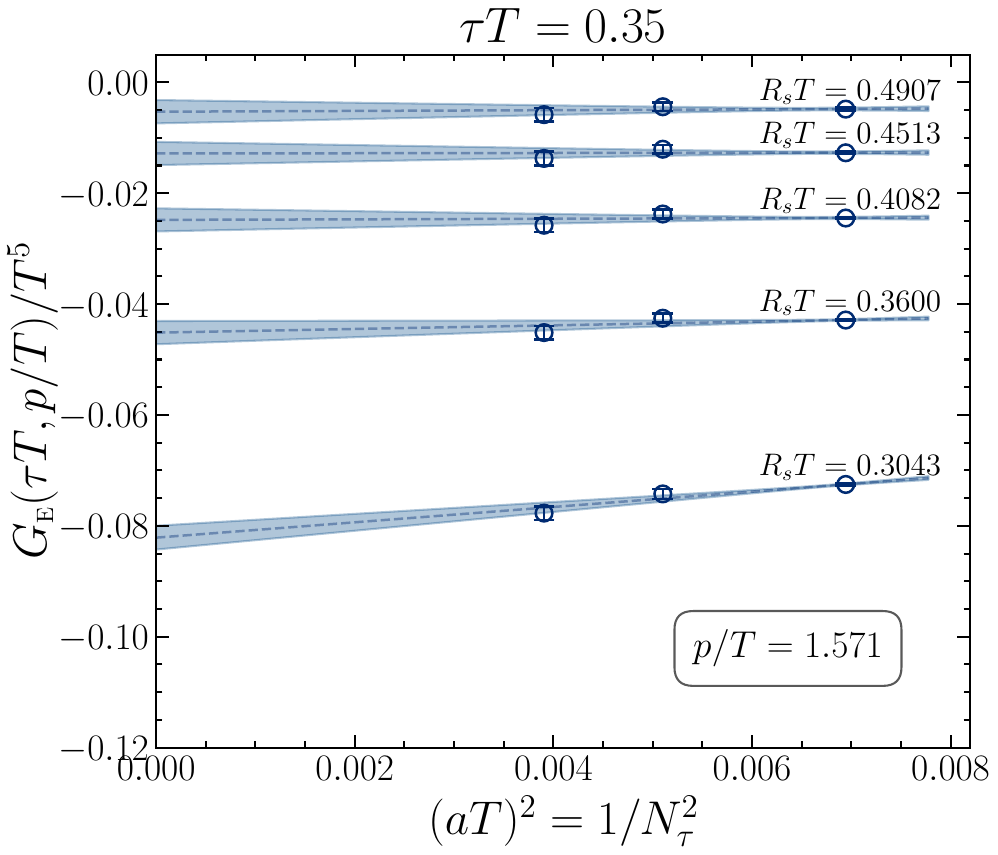}
\includegraphics[scale=0.48]{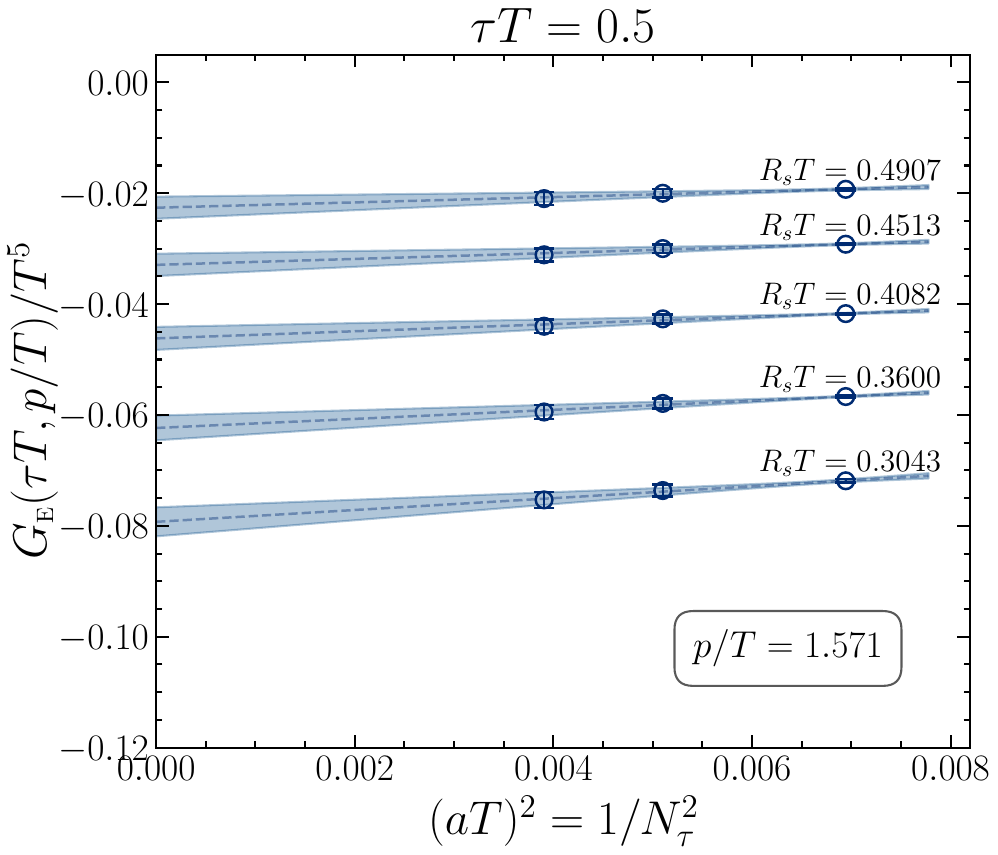}
\includegraphics[scale=0.48]{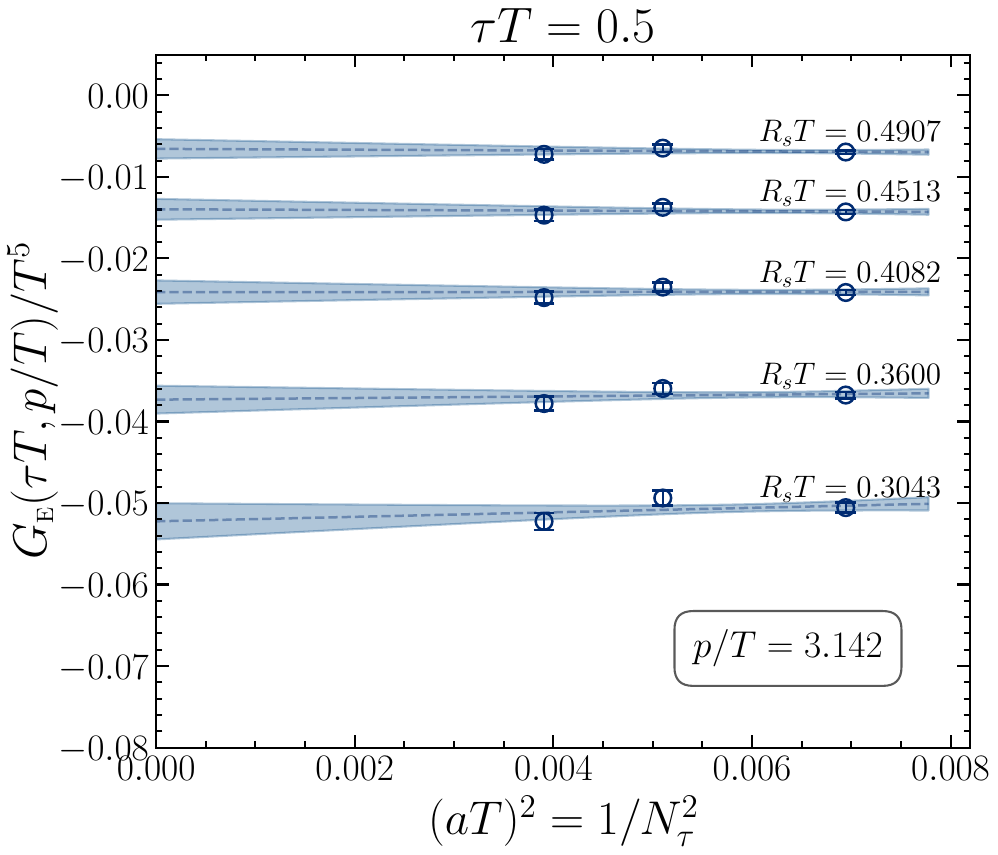}
\caption{Continuum extrapolation of $G_\E/T^5$ for two choices of $\tau T$ and $p/T$. The continuum limit is taken at fixed smoothing radius in physical units assuming linear corrections in $(aT)^2$, and is shown for several values of $\Rs T$.}
\label{fig:corr_cont_lim}
\end{figure}

\begin{figure}[!t]
\centering
\includegraphics[scale=0.48]{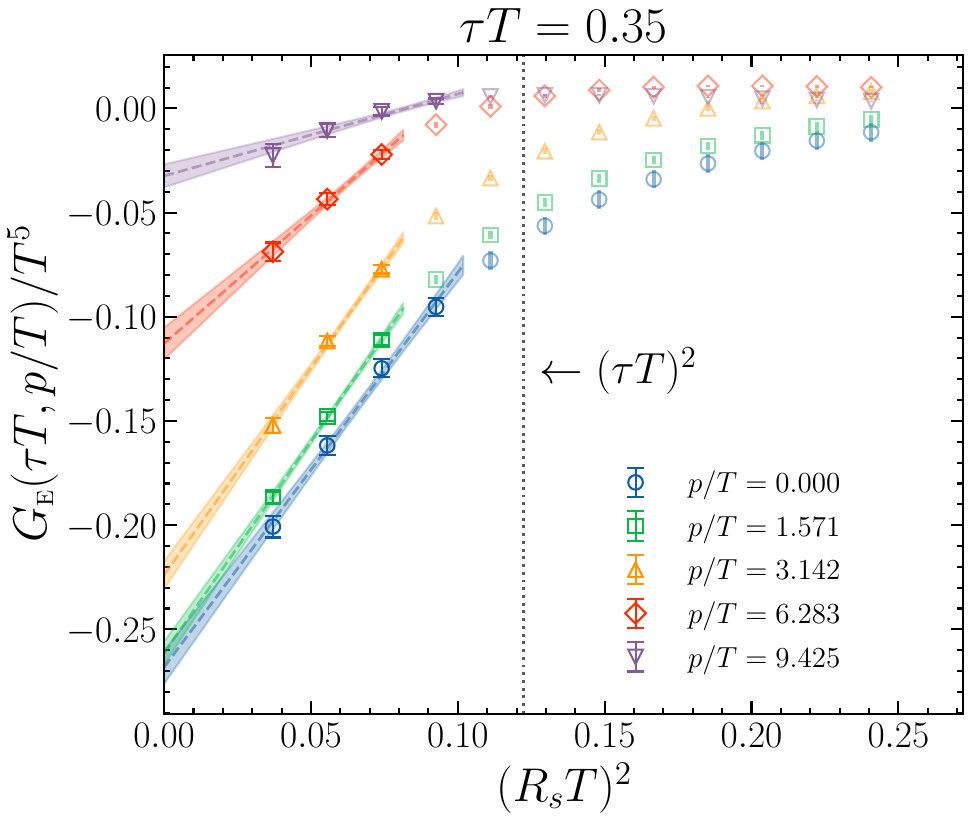}
\includegraphics[scale=0.48]{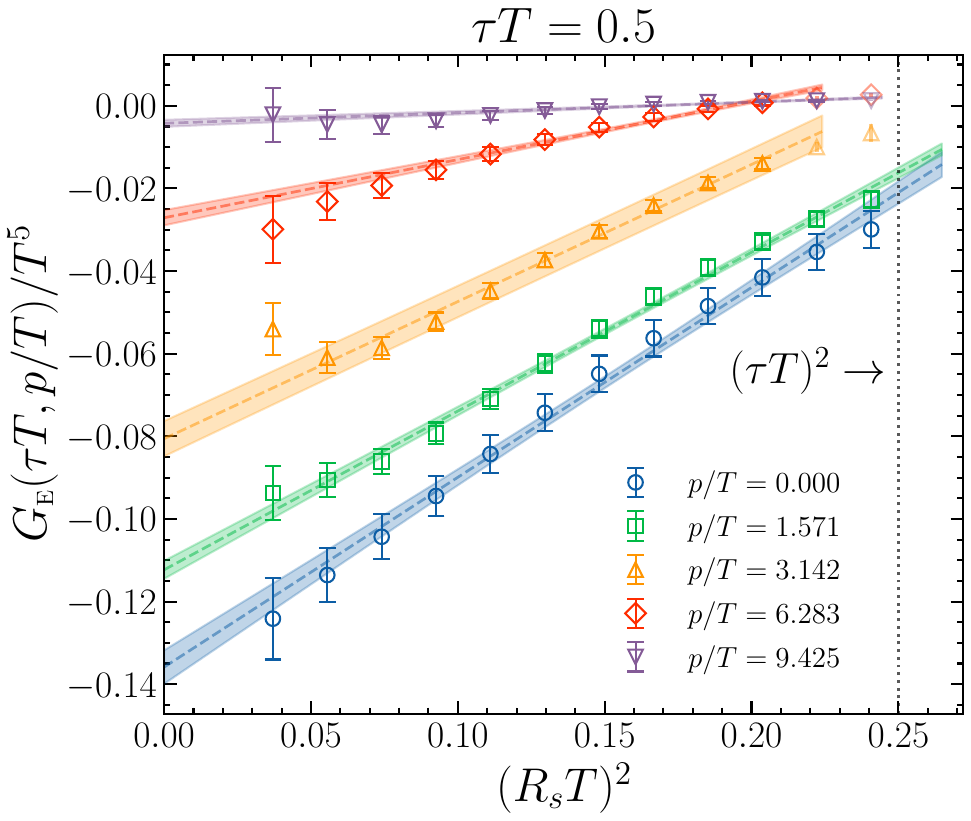}
\caption{Zero-cooling extrapolation of $G_\E/T^5$ for two choices of $\tau T$, and for several values of $p/T$. The extrapolation is performed assuming linear corrections in $(\Rs T)^2$, and the upper bound of the fit range was always chosen to ensure $\Rs<\tau$.}
\label{fig:corr_zerocool_lim}
\end{figure}

We will start illustrating the extraction of the topological rate from the double-extrapolated correlator (Method 1). The double extrapolation of $G_{\E}/T^5$ is always done taking first the continuum limit at fixed $\tau T$, $p/T$ and $\Rs T$ assuming leading quadratic corrections in $aT=1/N_\tau$,
\bee
\frac{1}{T^5}G_{\E}(\tau,p;a,\Rs) = \frac{1}{T^5}G_{\E}(\tau,p;\Rs) + C_1(\tau,p;\Rs) (aT)^2,
\eee
and then performing a zero-smoothing-radius limit assuming quadratic corrections in $\Rs T$:
\bee
\frac{1}{T^5}G_{\E}(\tau,p;\Rs) = \frac{1}{T^5}G_{\E}(\tau,p) + C_2(\tau, p) (\Rs T)^2.
\eee
The latter fit function is based on a general theoretical result mathematically proved using the gradient flow~\cite{Ce:2015qha,Ce:2016awn} (see also~\cite{Kotov:2018aaa,Altenkort:2020axj}), stating that $G_\E$ defined via positive-flow-time topological charge density sources receives linear corrections in $t_{\sst{\rm GF}}$ in the small-flow-time limit $t_{\sst{\rm GF}}\to 0$. By virtue of the smoothing radius matching in Eq.~\eqref{eq:cooling_gradientflow_matching} $t_{\sst{\rm GF}}\propto\Rs^2\propto\ncool$, thus one expects (and actually numerically finds) this result to hold when using cooling too~\cite{Bonanno:2023ljc,Bonanno:2023ple}.

\begin{figure}[!t]
\centering
\includegraphics[scale=0.42]{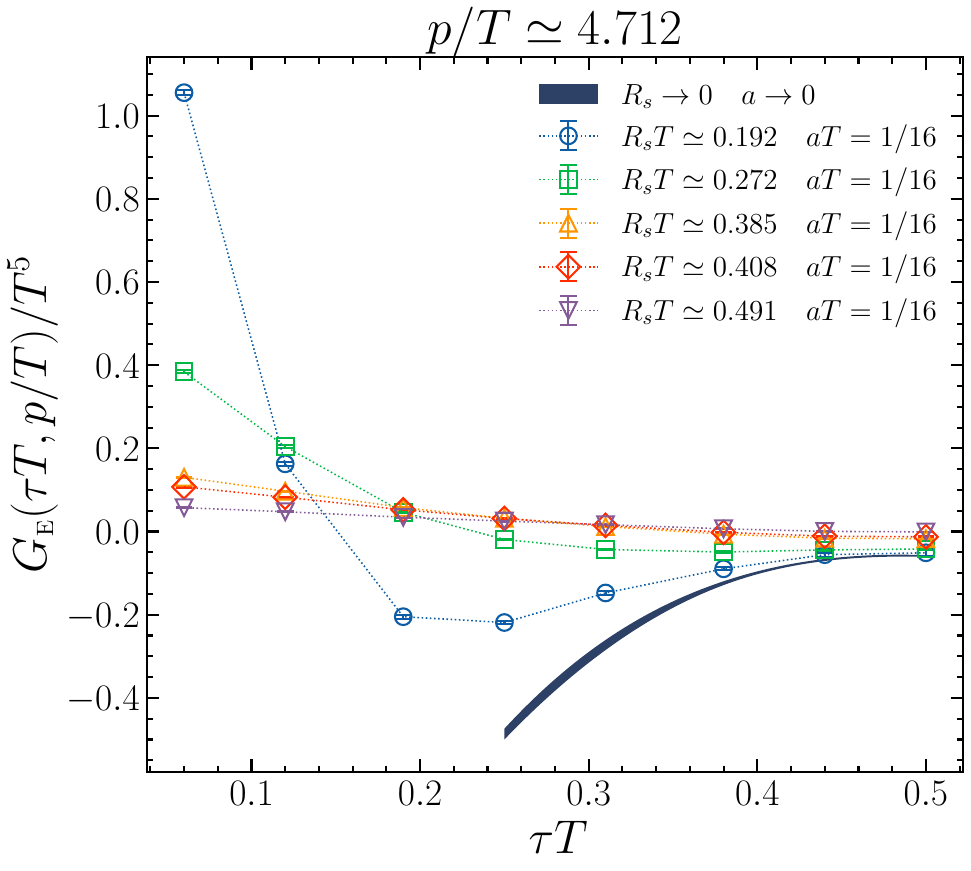}
\caption{Time-dependence of the correlators obtained at the finest lattice spacing for a few values of the smoothing radius, compared with the one of the double-extrapolated correlator. The plot refers to a fixed momentum $p/T$.}
\label{fig:corr_double_extr_vs_smooth_corrs}
\end{figure}

\begin{figure}[!t]
\centering
\includegraphics[scale=0.42]{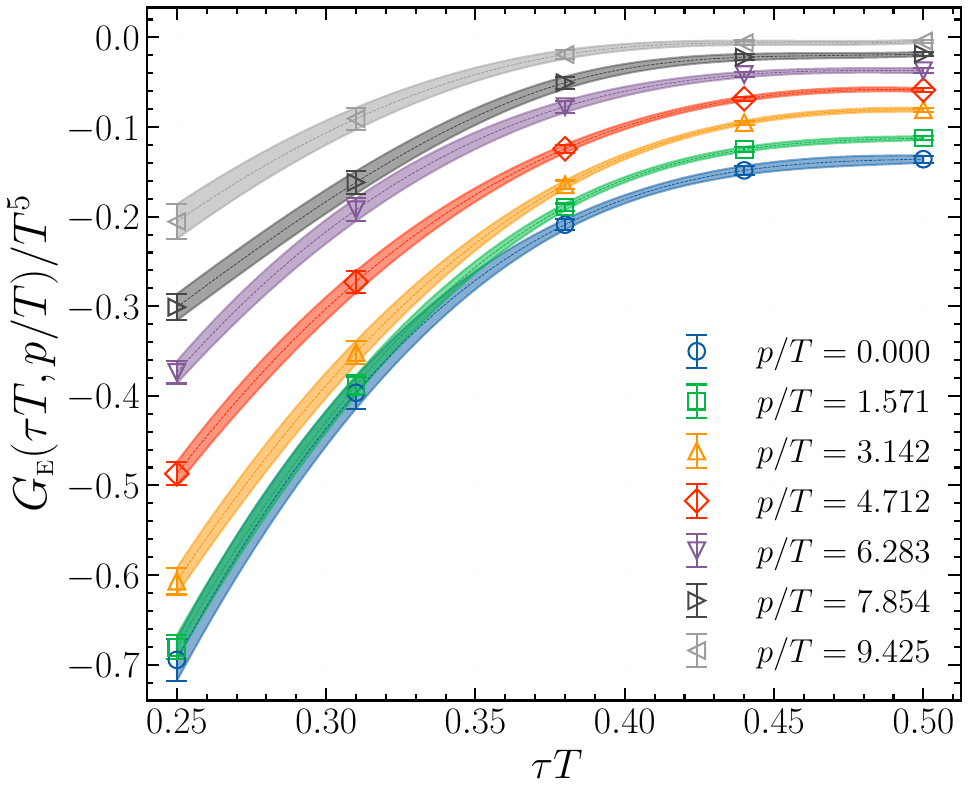}
\caption{Final results for the double-extrapolated correlator $G_\E/T^5$ as a function of $\tau T$ and $p/T$.}
\label{fig:corr_double_extr_vs_p}
\end{figure}

\begin{figure}[!htb]
\centering
\includegraphics[scale=0.4]{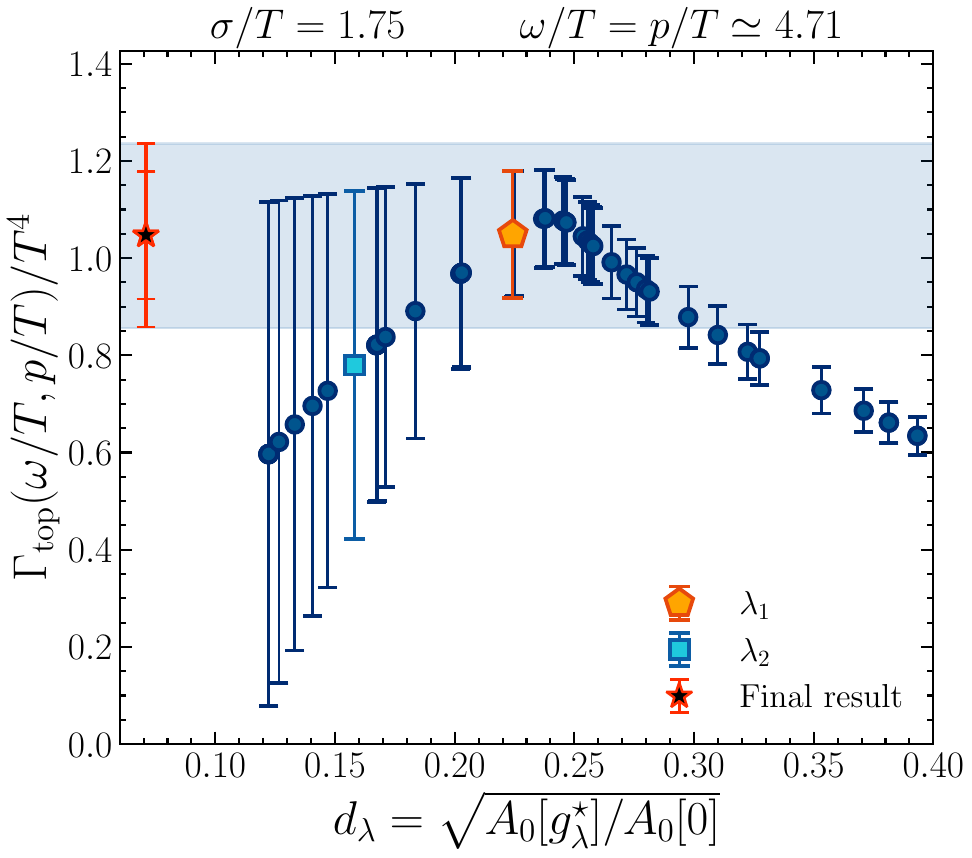}
\includegraphics[scale=0.4]{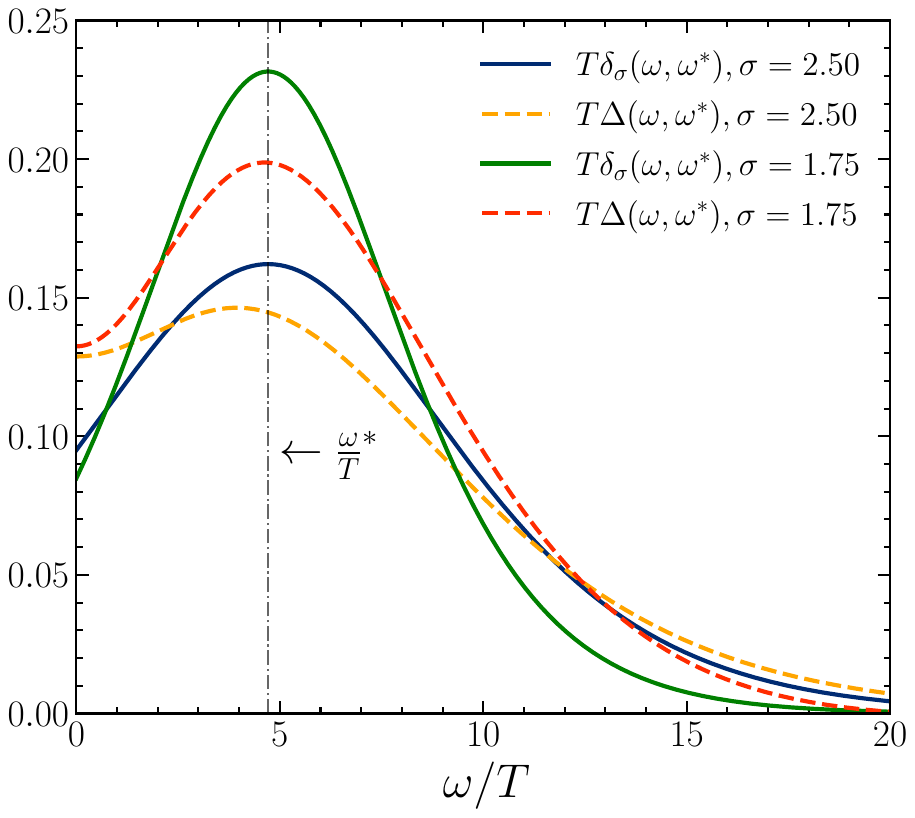}
\caption{Top panel: stability analysis on $\Gamma_\top$ for $\omegastar/T=p/T\simeq4.71$, and a smearing width $\sigma/T=1.75$ of the target kernel. The double error bars on the final result represent, respectively, the statistical and the quadrature sum of the statistical and systematic errors. These are assessed from $\Gamma_\top$ at, respectively, $\lambda=\lambda_1,\lambda_2$. Bottom panel: comparison of target and reconstructed smearing kernels for $\lambda=\lambda_1$ and $\sigma/T=1.75,2.5$.}
\label{fig:reco_ex_double_extr_corr}
\end{figure}

As already explained, while keeping fixed $p/T$ across ensembles with different lattice spacing can be done without any difficulty, each ensemble will possess different values of $\tau T$. Thus, to take the double limit at fixed time separation in physical units we interpolated the correlators at coarser lattice spacings to the time separations of the finest one via a cubic spline.

Another important aspect of the double limit is that one must adjust the range of smoothing radii employed for the extrapolation to ensure that $\Rs T< \tau T$. On top of this, we also chose the fit range so as to obtain reasonable values for the chi-squared. Given that the minimum smoothing radius accessible is fixed by our choice of ensembles, the condition $\Rs<\tau$ implies that the range of usable smoothing radii shrinks as $\tau T$ is reduced. This in practice puts a bound on the range of $\tau T$ for which a meaningful double limit can be performed: in the end, we could reliably compute a double extrapolation for $0.25 \le \tau T \le 0.5$.

Examples of the continuum limit and of the zero-smoothing-radius limit are reported, respectively, in Figs.~\ref{fig:corr_cont_lim} and~\ref{fig:corr_zerocool_lim}. The procedure is illustrated for two time separations, $\tau T=0.35, 0.5$, and for two values of the momentum, $p/T\simeq 1.571,3.142$. As it can be seen, the continuum limit turns out to be particularly smooth and almost flat, with a slope that is very mildly dependent on $\tau T$, $\Rs T$ or $p/T$.

Concerning the zero-smoothing radius limit, we observe that its slope become steeper as $\tau T$ is decreased at fixed momentum, while it flattens as $p/T$ is increased at fixed time separation. The first fact is related to the distortion introduced by smoothing on the correlation, which at fixed $\Rs$ becomes stronger as $\tau$ is decreased. This is illustrated in Fig.~\ref{fig:corr_double_extr_vs_smooth_corrs}, where the correlators at the finest lattice spacing for a few values of $\Rs$ are compared with the double-extrapolated one. The flattening of the $\Rs$ dependence as $p$ is increased is instead related to a general suppression of $G_\E$ as a function of the momentum at fixed time separation, as it can be seen from Fig.~\ref{fig:corr_double_extr_vs_p}, where the time-dependence of the final double-extrapolated correlators is displayed for several momenta up to $p/T\sim 10$, corresponding to $p/T=\frac{2\pi}{LT}k= \frac{\pi}{2}k$ with $k=0,1,2,\dots,6$, cf.~Eq.~\eqref{eq:discr_mom}.

\begin{figure}[!t]
\centering
\includegraphics[scale=0.44]{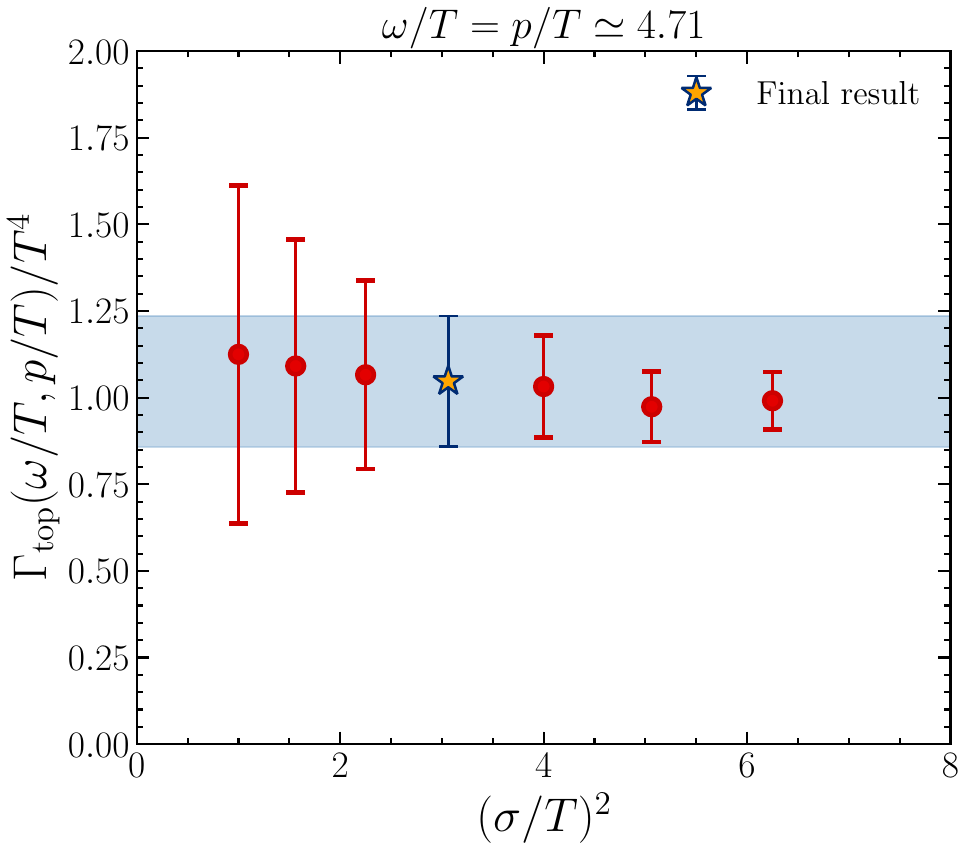}
\caption{Dependence of the topological rate on the squared smearing width $(\sigma/T)^2$.}
\label{fig:sigma_double_extr_corr}
\end{figure}

\begin{figure}[!t]
\centering
\includegraphics[scale=0.465]{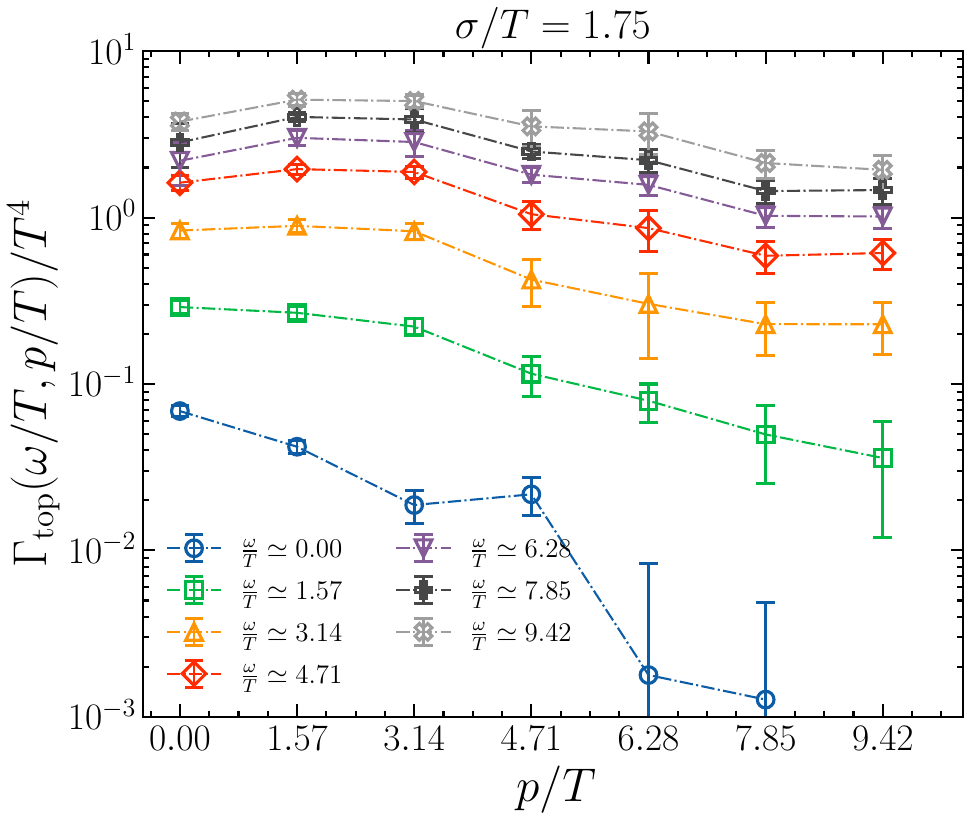}
\caption{Final results for the topological rate $\Gamma_\top$ as a function of the spatial momentum $p/T$ and of the energy $\omega/T$. Points connected with a dashed line refer to results obtained at the same value of $\omega/T$ and different spatial momenta.}
\label{fig:rate_from_double_extr_corr}
\end{figure}

We are now ready to apply the HLT method to the final double-extrapolated correlators. When performing the inversion, one has to select the value of the energy $\omega=\omegastar$ for which the calculation of the spectral density is performed, which can in general be fixed independently from the momentum $p$ (decided instead by the specific correlator entering the HLT procedure). Although for axion physics one is eventually interested only in the case $\omega\simeq p$, exploring the full parameter space is convenient to better understand the behavior of $\Gamma_\top$. For this reason, for each value of $p/T$, we performed the inversion for a few values of $\omega/T$ ranging from 0 to 10. More precisely, in view of the ultimate goal of computing the on-shell topological rate, we choose $\omegastar/T = \frac{\pi}{2}k$ with $k=0,1,\dots,6$, so as to match all values of $p/T$ at our disposal.

An example of inversion is displayed in Fig.~\ref{fig:reco_ex_double_extr_corr}. The reconstructed kernel (bottom panel) displays a similar behavior compared to the target one, signaling that the reconstruction quality is good. This is also shown in the stability analysis plot (top panel), which exhibits a plateau as a function of the figure of merit $d_\lambda$ as soon as $d_\lambda < 0.3$. In the stability analysis plot we have highlighted the two values $\lambda_1$ and $\lambda_2$ used to assess the statistical and systematic error on our final result for the topological rate, as well as the final result, where the two errors are summed in quadrature. The inversion was performed, for any given pair $(\omega/T,p/T)$, for several values of $\sigma/T\in[1.25,2.5]$ in steps of 0.25 in order to assess the possible systematic effects. The dependence on $\sigma/T$ turns out in the end to be rather mild, as shown in Fig.~\ref{fig:sigma_double_extr_corr}. This is in agreement with the general theoretical expectation that, in our case, leading finite-smearing-width correction should be of the order of $\mathcal{O}(\sigma^2)$. Given that practically no dependence on $\sigma/T$ can be observed within errors, the value $\sigma/T=1.75$, lying in the middle of the explored range, will be used.

The collection of the final results for $\Gamma_\top(\omega,p)$ are shown in Fig.~\ref{fig:rate_from_double_extr_corr}. In that figure we display several lines of points, each corresponding to the $p$ dependence of $\Gamma_\top$ at fixed value of $\omega$. As it can be seen, the rate at fixed $p$ grows as a function of the energy, while at fixed $\omega$ the rate seems to be almost constant for $p\lesssim \omega$, and then decreases above this threshold. Such decrease is more evident for $\omega/T\lesssim 3$, and becomes slower for larger energies, leading to an almost saturation of $\Gamma_\top$ above $p/T\gtrsim6$ for $\omega/T\gtrsim3$.

These observation are in general qualitative agreement with the behavior of $G_\E(\tau,p)$ observed in Fig.~\ref{fig:corr_double_extr_vs_p}. Indeed, at fixed momentum, increasing the energy means that one is probing the correlator at shorter time separations, where $G_\E$ exhibits a steep growth with respect to the behavior close to $\tau T = 0.5$, which dominates the $\omega=0$ case, implying the observed growth of $\Gamma_\top(\omega,p)$ as a function of $\omega$ at fixed $p$. On the other hand, at fixed time separation, the correlator $G_\E(\tau,p)$ is suppressed as a function of $p$, thus explaining the decrease of $\Gamma_\top(\omega,p)$ at large $p$ at fixed $\omega$.

\subsection{Double extrapolation of the topological rate (Method 2)}\label{sec:rate_calc_2}

\begin{figure}[!t]
\centering
\includegraphics[scale=0.43]{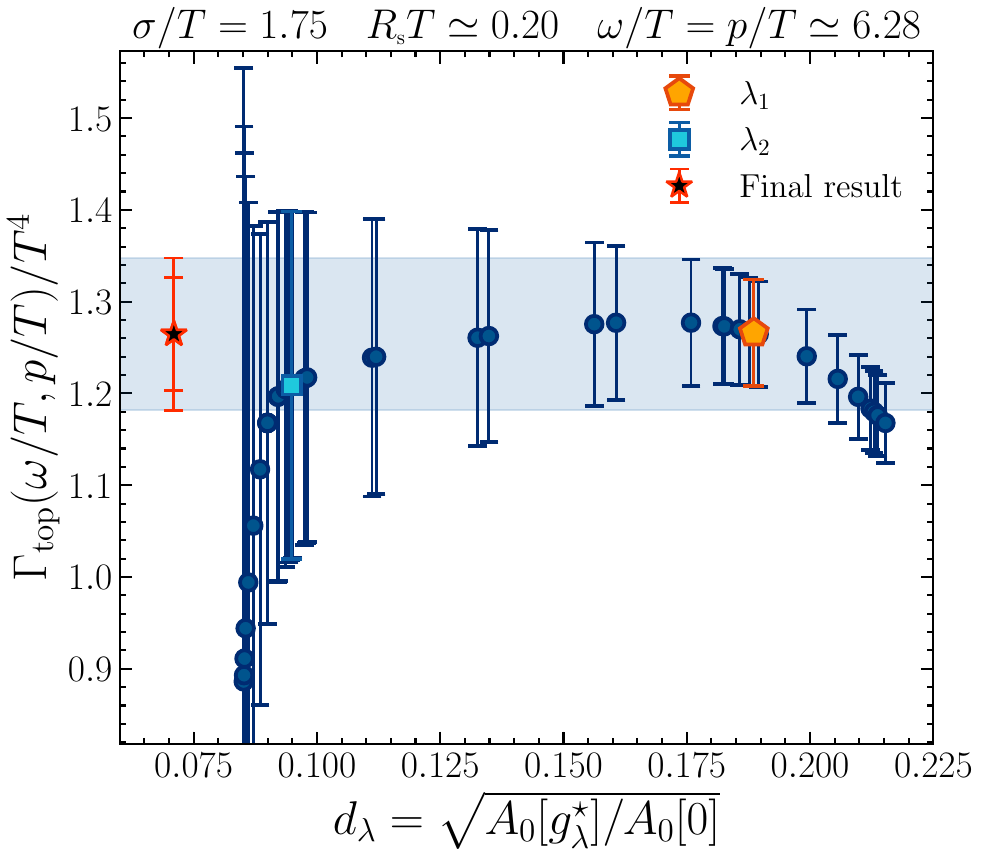}
\includegraphics[scale=0.43]{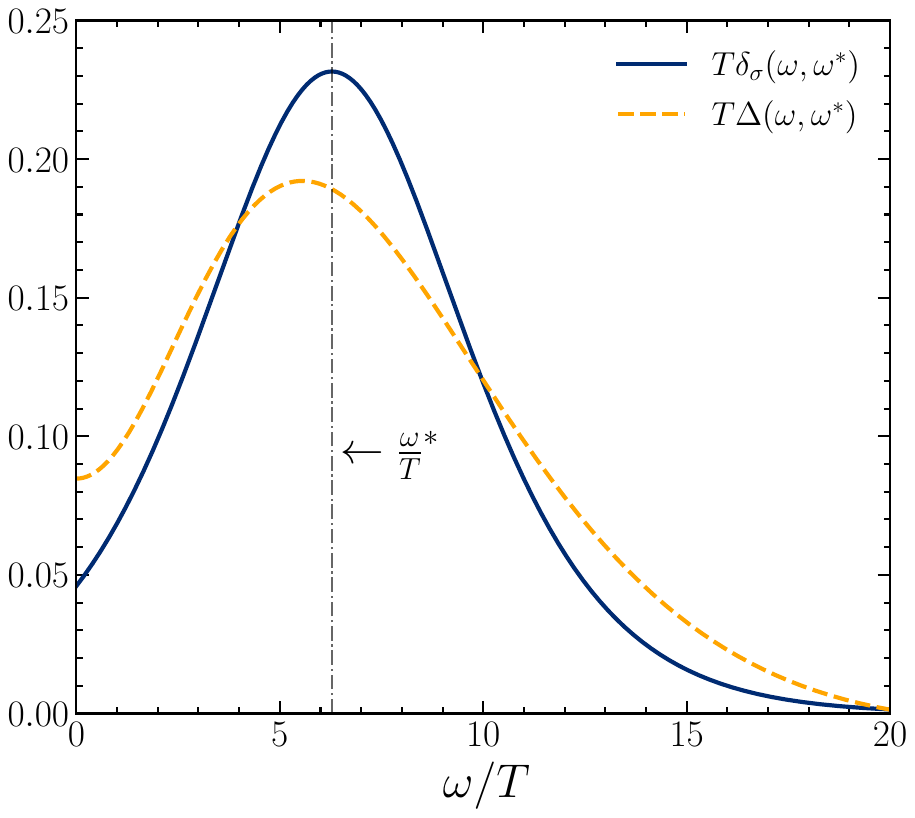}
\caption{Same as Fig.~\ref{fig:reco_ex_double_extr_corr} but in the case of a finite lattice spacing and finite smoothing radius correlator. The plot refers to $\omegastar/T=p/T\simeq6.28$, $\sigma/T=1.75$, $N_\tau=16$, $\Rs T\simeq0.20$.}
\label{fig:finite_stab_analysis}
\end{figure}

\begin{figure}[!t]
\centering
\includegraphics[scale=0.45]{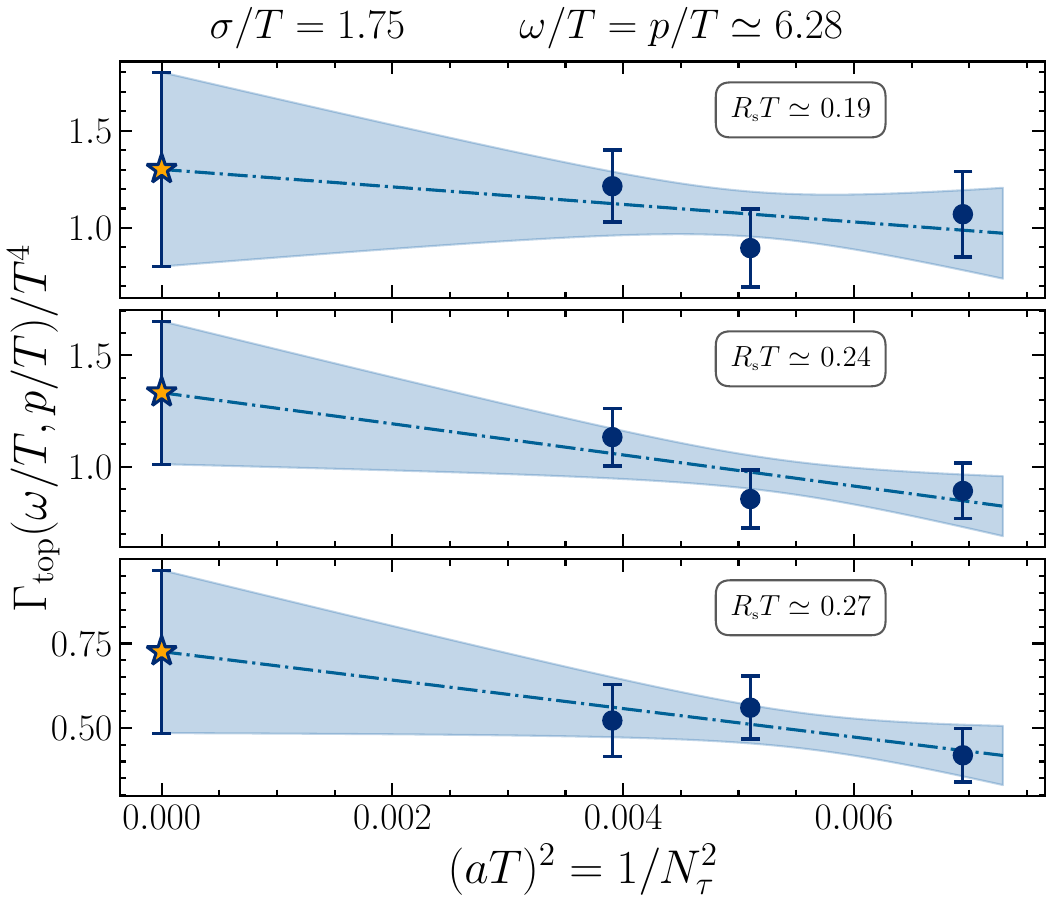}
\includegraphics[scale=0.39]{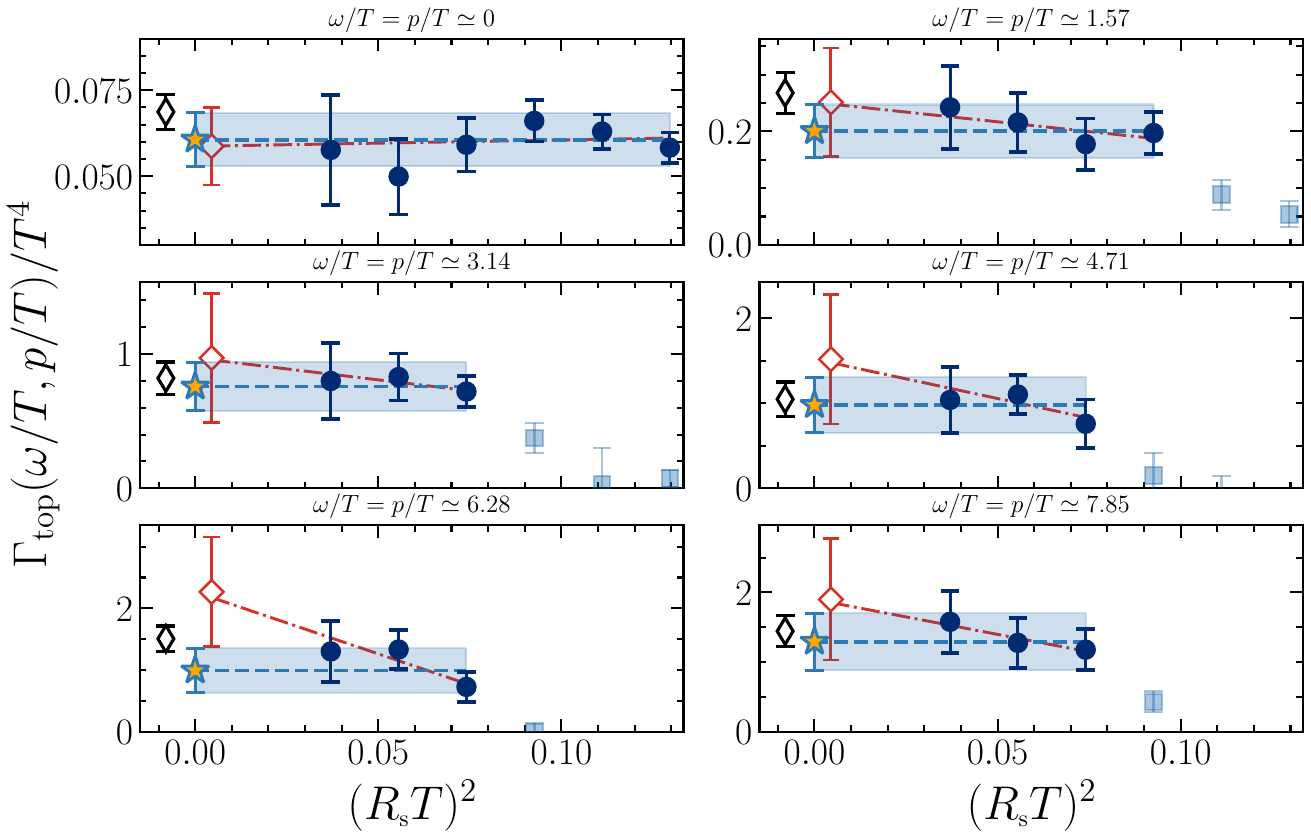}
\includegraphics[scale=0.45]{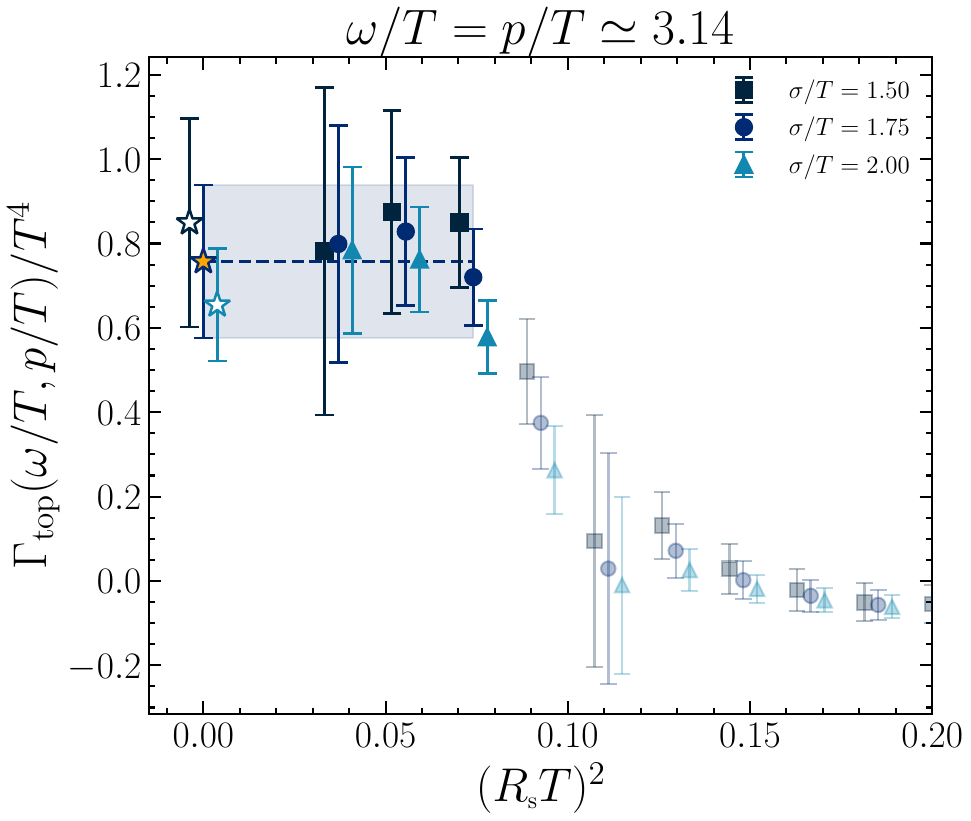}
\caption{Top panel: Continuum limit of the topological rate, extracted from finite lattice spacing and finite smoothing radius correlators, extracted from finite lattice spacing and finite smoothing radius correlators. Central panel: zero-smoothing radius limits of the continuum extrapolated determinations of the topological rate (filled points). These are compared with the topological rate obtained from the double-extrapolated correlators (empty points). Bottom panel: zero-smoothing limit extrapolation for a fixed value of $\omegastar$ and $p$ for a few choices of the smearing width $\sigma/T$.}
\label{fig:finite_doublelim}
\end{figure}

We will now cross-check the results obtained in Sec.~\ref{sec:rate_calc_1} (Method 1) by reversing the logic followed in that section. We will now first invert finite-$N_\tau$ and finite-$\Rs$ correlators via HTL, postponing the double continuum/zero-smoothing-radius limits directly on the topological rate (Method 2). Given that this procedure requires a much larger number of inversions, we will limit in this case to the on-shell case $\omega=p$, which is eventually the one that is interesting for axion physics applications.

An example of smearing kernel reconstruction, with its associated stability analysis study, is shown in Fig.~\ref{fig:finite_stab_analysis}. Also in this case, we observe the onset of a statistically-dominated plateau for $d_\lambda \lesssim 0.2$, signaling a good reconstruction with a characteristic width of the order of $\sigma$ around the desired energy $\omegastar$.

The double limit extrapolation of the topological rate obtained from finite-$N_\tau$ and finite-$\Rs$ correlators is shown in Fig.~\ref{fig:finite_doublelim} (top and central panels). First of all, we observe that the continuum limit is pretty smooth and presents a slope which is practically independent of $\Rs T$ (this behavior is observed for all energy values).

\begin{figure*}[!t]
\centering
\includegraphics[scale=0.37]{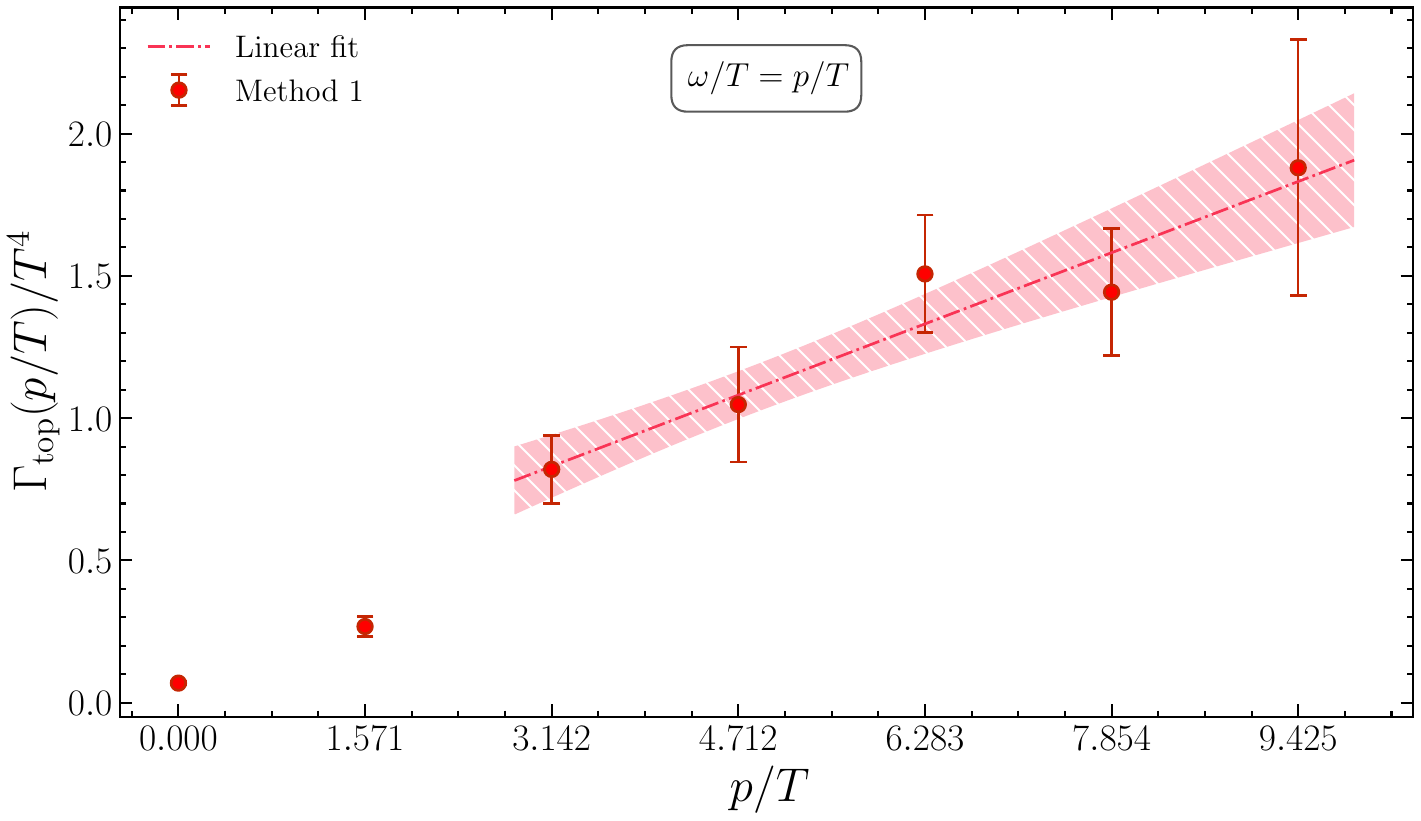}
\includegraphics[scale=0.37]{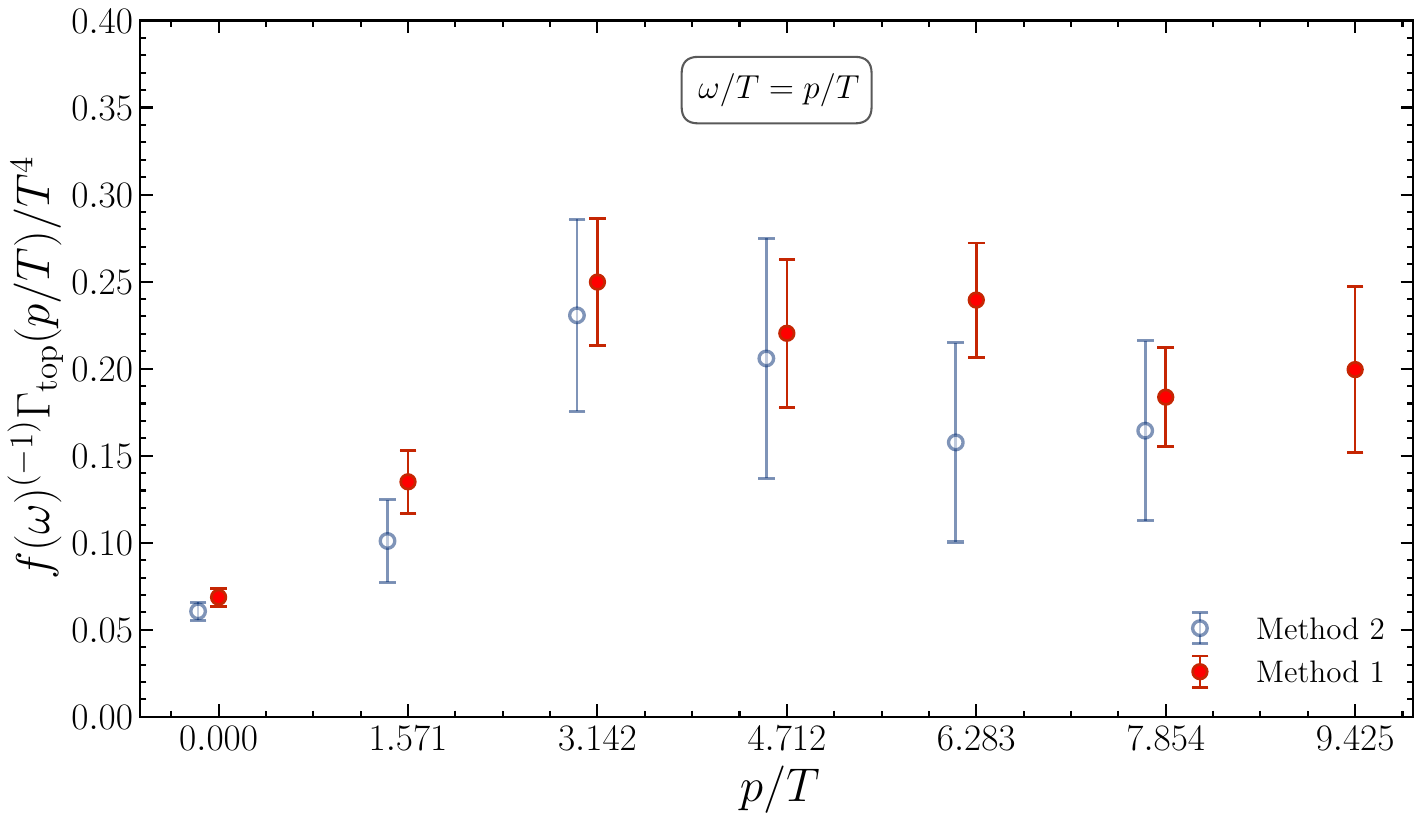}
\caption{Final results for the on-shell topological rate $\Gamma_{\top}(p/T)/T^4$ as a function of the momentum $p/T$ obtained from the two strategies described in Secs.~\ref{sec:rate_calc_1} (Method 1) and~\ref{sec:rate_calc_2} (Method 2). Right panel: comparison of the two methods. For clarity, data have been divided for the factor $f(\omega)=(\omega/T)/[1-\exp(-\omega/T)]$ Left panel: linear best fit of Method 1 data for $p/T\ge 2$.}
\label{fig:final_fig}
\end{figure*}

Then, we observe that, for small enough smoothing radii, the topological rate exhibits a plateau as a function of $\Rs T$. This behavior was already observed for the sphaleron rate in Refs.~\cite{Bonanno:2023ljc,Bonanno:2023thi}, and it is interpreted as an effective separation between the relevant time separations contributing to the topological rate and the UV cut-off set by $\Rs$. This is further confirmed by the fact that the plateau range reduces as $\omega/T$ is increased, in accord with our interpretation.

The zero-smoothing radius extrapolation was performed assuming a constant fit function, and a linear fit function in $(\Rs T)^2$, both giving perfectly agreeing results among themselves, and with the results obtained in Sec.~\ref{sec:rate_calc_1}. Moreover, we also observe a very good stability of the obtained results for the topological rate against the variation of $\sigma/T$ within the range already probed in the previous section, cf.~Fig.~\ref{fig:finite_doublelim} (bottom panel), further corroborating our findings.

\subsection{Discussion of the on-shell momentum-dependence of the topological rate}

The final results for the on-shell topological rate
\beq
\Gamma_{\top}(p) \equiv \Gamma_\top(\omega=p, p)
\eeq
obtained from the two discussed methods are reported in Tab.~\ref{tab:final_res}, and are displayed in Fig.~\ref{fig:final_fig} (right panel). We observe good agreement among the two methods, with Method 1 yielding more precise results. The rest of this section will be devoted to discuss the momentum dependence of $\Gamma_\top$. Method 1 numerical data will be used to this end.

Using Refs.~\cite{Masso:2002np,Graf:2010tv} to estimate the axion rate at asymptotically high temperatures from axion-gluon scattering leads to the following prediction for the momentum-dependence of the on-shell topological rate:
\beq
\Gamma_\top(p) = p (\alphas T)^3 \left\{A \log \left(\frac{pT}{\mD^2}\right) + B \right\},
\eeq
with $\alphas=g^2/(4\pi)$ the strong coupling constant and
\beq
\mD^2 = \frac{2N_c + N_{\sst{\rm f}}}{6} 4 \pi \alphas T^2
\eeq
the Debye mass~\cite{Gross:1980br} ($N_c$ and $N_{\sst{\rm f}}$ are, respectively, the number of colors and flavors). This calculation holds for asymptotically high temperatures $T\gg \mD\gg T_c$, and for asymptotically large momenta $p\gg\alphas T$. The quantities $A$ and $B$ are $\mathcal{O}(1)$ temperature-independent constants, but are expected to grow as a function of the number of flavors, leading to an enhancement of the topological rate in the presence of light quark masses, similarly to what has been observed on the lattice in the zero-momentum case~\cite{Bonanno:2023ljc,Bonanno:2023thi}. Thus, $\Gamma_\top$ grows linearly with $p$ up to parametrically small logarithms.

\begin{table}[!t]
\centering
\begin{tabular}{|c|c|c|c|}
\hline
&&&\\
$\,\, k \,\,$ & $\,\,\dfrac{p}{T} = \dfrac{2\pi}{LT}k\,\,$ & \makecell{$\,\,\dfrac{1}{T^4}\Gamma_{\top}\left(\dfrac{p}{T}\right)\,\,$\\ \\(Method 1)} & \makecell{$\,\,\dfrac{1}{T^4}\Gamma_{\top}\left(\dfrac{p}{T}\right)\,\,$\\ \\(Method 2)}\\
&&&\\
\hline
0 & 0    & 0.0687(52) & 0.0606(78) \\
1 & 1.57 & 0.268(36)  & 0.200(47)  \\
2 & 3.14 & 0.82(12)   & 0.757(181) \\
3 & 4.71 & 1.05(20)   & 0.98(33)   \\
4 & 6.28 & 1.51(21)   & 0.99(36)   \\
5 & 7.85 & 1.44(22)   & 1.29(41)   \\
6 & 9.42 & 1.87(45)   &            \\
\hline
\end{tabular}
\caption{Final results for the on-shell ($\omega=p$) topological rate $\frac{1}{T^4}\Gamma_{\top}(p/T)$ for $T/T_c\simeq 1.24$ and $LT=4$ with the two methods described in Secs.~\ref{sec:rate_calc_1} (Method 1) and~\ref{sec:rate_calc_2} (Method 2). For method 2, we could not provide a result with Method 2 due to $p/T\sim 9.4$ being close to the cut-off scale $\Lambda/T=1/(aT)=N_\tau$ for the coarsest lattice spacing $N_\tau=12$. This fact prevented us from obtaining a reliable extraction of the spectral density in that case, and thus the possibility of taking the continuum limit.}
\label{tab:final_res}
\end{table}

Using $N_c=3$ and $N_{\sst{\rm f}}=0$ and the full expression of the Debye mass, one obtains the following:
\beq\label{eq:semiclassic}
\frac{1}{T^4}\Gamma_{\top}\left(\frac{p}{T}\right) = \alphas^3 \frac{p}{T} \left\{A \log \left(\frac{1}{4\pi\alphas}\frac{p}{T}\right) + B \right\}.
\eeq
Clearly, one does not expect Eq.~\eqref{eq:semiclassic} to hold in the non-perturbative regime we are exploring, close to $T_c$. Nonetheless, we observe that a linear growth of $\Gamma_\top(p)$ for $p\ge0$ seems to describe reasonably well our data. Indeed, assuming a first-order Taylor-expansion fit ansatz of the type:
\beq
\frac{1}{T^4}\Gamma_\top\left(\frac{p}{T}\right) = c_0 + c_1 \frac{p}{T} + \mathcal{O}\left(\frac{p^2}{T^2}\right),
\eeq
we find a good fit quality and a stable result for the slope coefficient $c_1$ if the fit range is varied excluding the smallest momenta. As an example, we find:
\beq
c_1 &= 0.180(13), \qquad p/T \ge 0,\\
c_1 &= 0.226(24), \qquad p/T \ge 1,\\
c_1 &= 0.158(43), \qquad p/T \ge 2.
\eeq
The reduced chi-squared read, respectively, 10.2/5, 4.75/4 and 1.17/3. The most conservative fit in the range $p/T \ge 2$ is shown in Fig.~\ref{fig:final_fig} (left panel). A best fit of all data with an ansatz of the type:
\beq
\frac{1}{T^4}\Gamma_\top\left(\frac{p}{T}\right) = c_0 + c_1 \left(\frac{p}{T}\right)^\alpha
\eeq
yields $\alpha=1.20(11)$ with a reduced chi-squared of 6.76/4. This result is again perfectly compatible with a linear growth of $\Gamma_\top$ as a function of $p$ in the explored range.

\section{Conclusions}\label{sec:conclu}

In this article we have presented a proof-of-concept study of the topological rate at non-zero momentum and energy in quenched QCD in the high temperature phase, focusing on a single temperature $T \simeq 1.24 \, T_c \approx 360$ MeV, and exploring momenta up to $p/T \sim 10$.

The topological rate was computed following the two methods put forward in Refs.~\cite{Bonanno:2023ljc,Bonanno:2023thi}, both relying on the HLT strategy to compute spectral densities from lattice correlators. One method consists in performing a double continuum and zero-smoothing limit on lattice correlators, and then to invert the resulting double-extrapolated two-point function. The second method consists instead in the inversion of finite-lattice-spacing and finite-smoothing-radius correlators, postponing the double limit on the topological rate itself. In both cases, we are able to obtain controlled continuum and zero-smoothing limits, as well as to employ sufficiently small smearing widths in the HLT inversion to be insensitive to finite-smearing width effects. Moreover, we find very good agreement among the two different strategies, confirming our good control over all possible sources of systematic effects. The choice of smoothing radii employed for the calculation of the topological rate has been justified by observing that the continuum limit of the topological susceptibility is well-independent of $\Rs$, and agreeing with previous determinations in the literature at the same temperature.

Focusing on the  on-shell case $\omega=p$, we find a linear increase of $\Gamma_\top$ as a function of $p$. Notably, this is also the expected semiclassical behavior for asymptotically-large momenta and temperatures. This result may seem at first glance in contradiction with the observed suppression with increasing momentum of the two-point function of the topological charge density at all time separations. However, we have shown that this is actually not the case. By virtue of the possibility of exploring the off-shell cases $\omega\ne p$, we have shown that $\Gamma_\top$ is suppressed with $p$ only at fixed $\omega$, while it grows at fixed $p$ as a function of $\omega$. This can be explained based on the fact that the Euclidean topological correlator $G_{\E}(\tau,p)$ grows and eventually diverges for $\tau \to 0^+$ at fixed $p$. Thus, when focusing on the case $\omega=p$, there are two competing effects. Given that the observed $p$-suppression at fixed $\omega$ is much milder than the growth with $\omega$ at fixed $p$, this in the end leads to a growth of the on-shell rate $\Gamma_\top(p)$ as a function of $p$.

In the next future, we plan to extend the study presented in this paper to the full QCD case with physical quark masses. Our goal is to explore if and how the growth with the momentum of the topological rate is enhanced with light quark masses, and how it changes as a function of the temperature.

\begin{figure*}[!htb]
\centering
\includegraphics[scale=0.39]{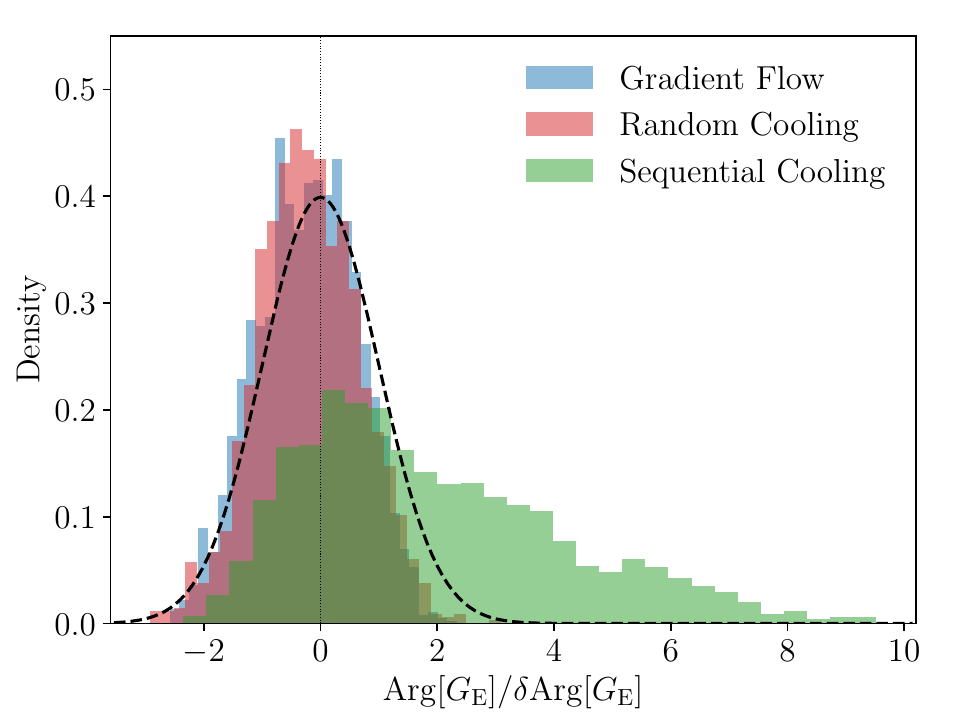}
\includegraphics[scale=0.37]{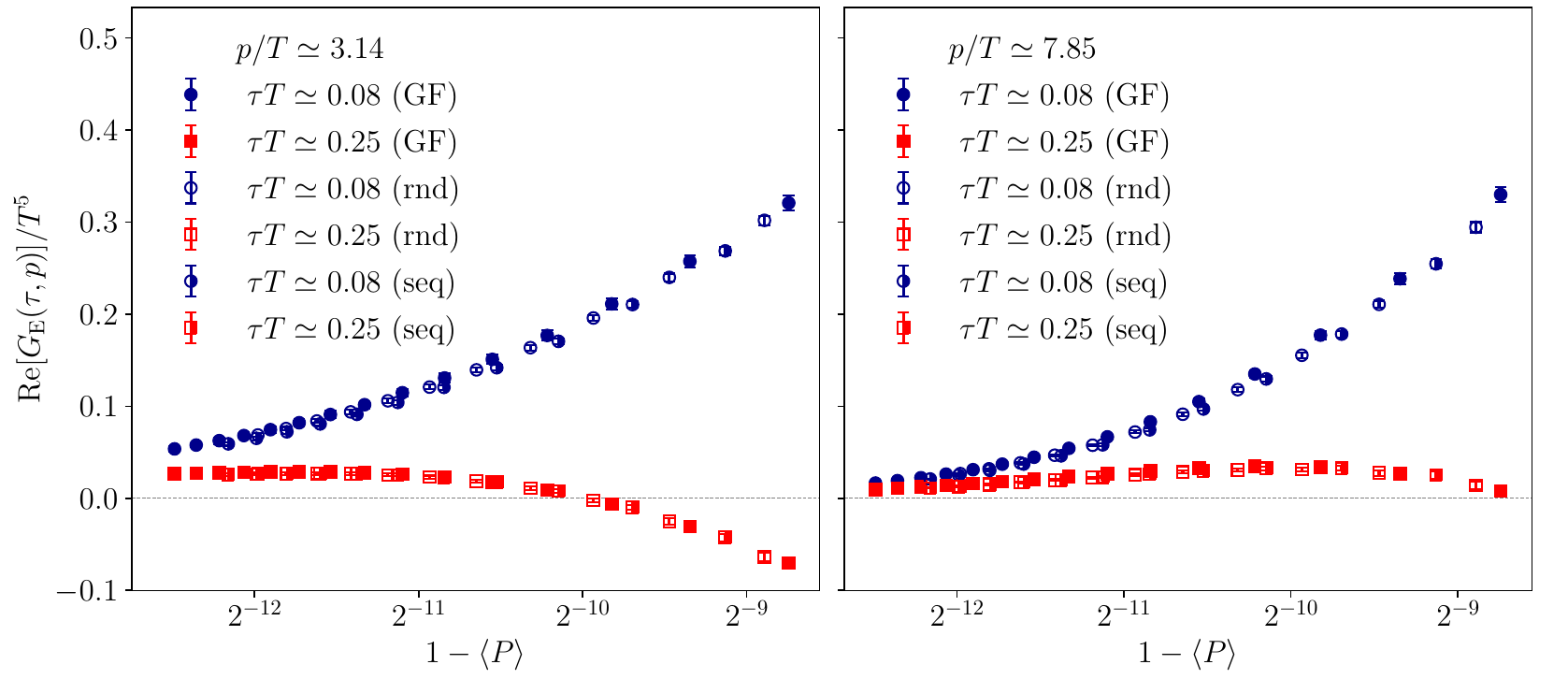}
\caption{Left panel: Histograms of the complex phase of $G_{\E}(\tau,p)$, normalized by its statistical uncertainty. Results refer to $N_\tau = 12$, $0 < ap < \pi/N_s$, $ 0 < \tau/a < N_\tau / 2$, and are obtained using gradient flow (GF) and the two implementations of cooling described in the text: random (rnd) and sequential (seq). For gradient flow and randomized cooling, histograms are compatible with a Gaussian distribution with zero mean and unit variance (black dashed line), implying a vanishing $\mathrm{Im}[G_{\E}(\tau,p)]$. This is not the case for sequential cooling, which explicitly breaks time-reversal symmetry. Central and right panels: comparison of the real part of $G_{\E}(\tau,p)$. All three smoothing methods give compatible results for $\mathrm{Re}[G_{\E}(\tau,p)]$ as a function of the average plaquette $\langle P \rangle$, here used as a proxy for the smoothing radius $R_s$.}
\label{fig:cooling_comp}
\end{figure*}

\acknowledgments
We thank N.~Bellini for collaboration in the initial stages of this project. It is a pleasure to thank M.~Gorghetto, P.~Lowdon, and A.~Notari for useful discussions. R.~Dionisio acknowledges funding from the European Union (EU) Next Generation EU (NGEU) -- National Recovery and Resilience Plan (NRRP) -- MISSION 4 COMPONENT 1, INVESTMENT N.4.1 -- CUP N.~I51J24000160007 (PhD Cycle XL, Ministerial Decree no.~629/2024). F.~Sanfilippo is supported by ICSC -- Centro Nazionale di Ricerca in High Performance Computing, Big Data and Quantum Computing,
funded by the EU NGEU -- and by the Italian Ministry of University and Research (MUR) project FIS 00001556. This work has also been supported by the project “Non-perturbative aspects of fundamental interactions, in the Standard Model and beyond” funded by MUR, Progetti di Ricerca di Rilevante Interesse Nazionale (PRIN), Bando 2022, grant 2022TJFCYB (CUP I53D23001440006). Numerical calculations have been performed on the \texttt{Leonardo} machine at Cineca, based on the agreement between INFN and Cineca, under projects INF25\_npqcd, INF26\_npqcd.

\section*{Data Availability Statement}

The data that support the findings of this article will be made available upon reasonable request.

\section*{Appendix}
\appendix
\section{Comparison among different cooling implementations}\label{app:cooling}

As discussed in Sec.~\ref{subsec:smoothing_role}, Wilson cooling smooths gauge configurations by aligning each link with the corresponding staple, minimizing the Wilson action locally. Since the smoothed configuration depends on the order in which gauge links are visited, different implementations of cooling \emph{may} give different results at finite smoothing radius. In the implementation used in this work, the space-time directions are swept in lexicographic order, and, for each fixed direction, gauge links at even lattice sites are updated before those at odd lattice sites. This ordering is computationally efficient because gauge links with same parity and direction can be updated in parallel, as they do not appear in the staple of one another, However, systematically cooling a sub-lattice of given parity before the other explicitly breaks time-reversal symmetry, since time reversal exchanges even and odd lattice sites. This is why the correlator of the topological charge density $G_{\E}(\tau, p)$ in \cref{eq:corr_lat_def} can have a non-vanishing imaginary part at finite smoothing radius.

To check that this does not affect the real part of $G_{\E}(\tau, p)$, we tested two other smoothing methods at our coarsest lattice spacing: gradient flow, which preserves time-reversal symmetry trivially, and an implementation of cooling that we call \emph{random cooling}, in contrast with the \emph{sequential cooling} discussed earlier. In random cooling, the symmetry is preserved by randomizing the updating order of space-time directions and site-parity for every configuration and at every cooling step. The outcome of this test is illustrated in \cref{fig:cooling_comp} (left panel), which shows that the complex phase $\mathrm{Arg}[G_{\E}(\tau, p)]$ is compatible with zero with gradient flow and random cooling, in contrast with sequential cooling (even though in this case the imaginary part is smaller than the real part by at least one order of magnitude in most cases). More importantly, we found that these three smoothing procedures give perfectly compatible results for $\mathrm{Re}[G_{\E}(\tau,p)]$ when compared at fixed expectation value of the plaquette $\langle P \rangle \equiv \frac{1}{18N_s^3 N_\tau}\sum_{\mu>\nu}\sum_x\braket{\mathrm{Re}\Tr[U_{\mu\nu}(x)]}$, as done in Ref.~\cite{Bonati:2014tqa}. Some of these comparisons are shown in \cref{fig:cooling_comp} (central and right panels). Thus, we conclude that the breaking of time-reversal symmetry of sequential cooling does not introduce any artifact in the calculation of the real part of the correlator at finite smoothing radius.

\end{document}